\documentclass[twocolumn,numberedappendix,appendixfloats]{openjournal}

\usepackage{xcolor}
\usepackage{textgreek}
\usepackage[utf8]{inputenc}
\usepackage[english]{babel}

\usepackage{hyperref}
\hypersetup{
    unicode, 
    colorlinks=true,
    linkcolor=linkcolor,
    citecolor=linkcolor,
    filecolor=linkcolor,
    urlcolor=linkcolor,
}
\usepackage{color,colortbl}
\definecolor{linkcolor}{rgb}{0.0,0.3,0.5}
\usepackage{tensind}
\tensordelimiter{?}
\DeclareGraphicsExtensions{.bmp,.png,.jpg,.pdf}
\usepackage{verbatim}
\usepackage[normalem]{ulem}
\usepackage{orcidlink}
\usepackage{soul}
\usepackage{amsmath}
\usepackage{tabularx}
\usepackage{makecell}

\graphicspath{ {./figs/} }

\defcitealias{Aihara2026}{Aihara et al., submitted}

\begin{document}
\title{
Kinematics of the Weak Cool-Core Cluster A3571 Observed with XRISM: Low Cooling Rate Balanced by Low Heating Rate?
}

\author{
Hannah McCall$^{1}$\orcidlink{0000-0003-3537-3491}, 
Irina Zhuravleva$^{1}$\orcidlink{0000-0001-7630-8085}, 
Kyoko Matsushita$^{2}$,
Annie Heinrich$^{1}$\orcidlink{0000-0002-7726-4202},
Congyao Zhang$^{3,1}$\orcidlink{0000-0001-5888-7052},
Eugene Churazov$^{4,5}$\orcidlink{0000-0002-0322-884X},
William Forman$^{6}$\orcidlink{0000-0002-9478-1682},
Ildar Khabibullin$^{7,4}$\orcidlink{0000-0003-3701-5882},
Kotaro Fukushima$^{2}$\orcidlink{0000-0001-8055-7113},
Daniele Rogantini$^{1}$\orcidlink{0000-0002-5359-9497},
Itsuki Aihara$^{2}$,\orcidlink{0009-0009-8380-1260}
Christine Jones$^{6}$\orcidlink{0000-0003-2206-4243},
Kazunori Suda$^{2}$\orcidlink{0009-0003-3920-131X}
}

\email{hannahmccall@uchicago.edu}
\affiliation{$^{1}$ Department of Astronomy \& Astrophysics, University of Chicago, 5640 S Ellis Ave., Chicago, IL 60637 USA}
\affiliation{$^{2}$Department of Physics, Tokyo University of Science, 1-3 Kagurazaka, Shinjuku-ku, Tokyo 162-8601, Japan}
\affiliation{$^{3}$Department of Theoretical Physics and Astrophysics, Faculty of Science, Masaryk University, 61137 Brno, Czech Republic}
\affiliation{$^{4}$Max Planck Institut f\"ur Astrophysik, Karl-Schwarzschildstr.~1, D-85748 Garching, Germany}
\affiliation{$^{5}$Space Research Institute (IKI), Profsoyuznaya 84/32, Moscow 117997, Russia}
\affiliation{$^{6}$Harvard-Smithsonian Center for Astrophysics, 60 Garden St., Cambridge, MA, 02138 USA}
\affiliation{$^{7}$Rudolf Peierls Centre for Theoretical Physics, Department of Physics, University of Oxford, Clarendon Laboratory, Parks Rd, Oxford, OX1 3PU, United Kingdom}

\begin{abstract}
Most XRISM galaxy cluster observations to date have focused on AGN feedback or actively merging systems. 
The weak cool-core cluster A3571 was observed in four XRISM Cycle 1 pointings, enabling the study of gas kinematics in a relaxed, AGN-feedback-free system. 
We present measurements of the velocity dispersion and bulk velocity in the core regions of A3571, out to $120$\,kpc. 
The velocity dispersion is relatively uniform across all regions ($\sim100-120 ~\mathrm{km~s^{-1}}$), except in the northern gas sloshing elongation, where a $68\%$ upper limit of $68~\mathrm{km~s^{-1}}$ is obtained.
The core Mach number and non-thermal pressure fraction of A3571 are lower than in the extremely relaxed cluster A2029 and below predictions from cosmological simulation suites. 
Despite relatively low velocity dispersion values, the derived turbulent heating rate (under the assumption of isotropic Kolmogorov turbulence) is sufficient to offset cooling losses within the considerable uncertainties in the studied regions. This suggests that sloshing motions may contribute significantly to the heating budget.
Comparing XRISM observations of merging and relaxed clusters, we find that mergers exhibit an average Mach number of $0.29\pm0.07$, nearly twice that of the relaxed sample, which is consistent with predictions from non-radiative cosmological simulations.
A3571 is a promising target for resonant scattering studies; however, simulations indicate that deeper observations are required to obtain reliable turbulent velocities via the $z/w$ line ratio.
\end{abstract}

\begin{keywords}
    {Galaxy clusters, Intracluster medium, High energy astrophysics, High resolution spectroscopy}
\end{keywords}

\maketitle

\section{Introduction}
\label{sec:intro}

For over two decades, X-ray CCD detectors such as Chandra, XMM-Newton, Suzaku, and eROSITA have transformed our understanding of galaxy clusters by providing spatially resolved observations of the hot intracluster medium (ICM). These missions enabled detailed studies of cluster thermodynamics \citep[for selected reviews, see][]{Boehringer2010,Thermo_profs2022}; however, the moderate spectral resolution of CCD detectors limited their ability to acquire direct ICM gas velocity measurements. The launch of XRISM \citep{XRISM_2025}, following the pioneering high-resolution spectroscopic measurements of Hitomi \citep{Hitomi2016,Hitomi2018}, now enables direct constraints on line-of-sight gas velocities both via emission-line shift and line broadening measurements, and in some systems via resonant scattering. These measurements provide a means of constraining turbulent and bulk gas motions in clusters, allowing investigations of active galactic nuclei (AGN)-driven motions, merger-induced turbulence and bulk flows, and sloshing in the ICM.

Many early XRISM cluster observations have targeted bright AGN feedback systems and actively merging clusters \citep[for a selected sample, see][]{XRISM_Centaurus,XRISM_Coma,XRISM_Perseus2026,M87_paper1}. However, the current sample of observed intermediate objects, including dynamically relaxed clusters with weak or absent AGN activity, remains comparatively limited. Such clusters are valuable because sloshing-induced motions (a long-lasting signature of merging activity) can be observed in isolation from the strong influence of ongoing AGN feedback.

\begingroup 
    \setlength{\tabcolsep}{10pt} 
    \renewcommand{\arraystretch}{1.5} 
    \setlength\extrarowheight{2pt}
    \begin{table*}
        \centering
        \begin{tabular}{ c c c c c }  
            ObsID & Starting date & R.A. & Dec. & Exposure time (ks)\\
            \hline \hline     
            201023010 & 01-09-2025 & 206.86602 & -32.8884 & 174.6  \\
            201025010 & 01-04-2025 & 206.92752 & -32.85464 & 72.4 \\
            201024010 & 01-06-2025 & 206.86576 & -32.83701 & 137.4 \\
            201095010 & 12-30-2024 & 206.86605 & -32.85329 & 192.1 \\
            \hline \hline  
        \end{tabular}  
        \caption{Summary of the XRISM/Resolve A3571 observations analyzed in this paper.}
        \label{tab:observations}
    \end{table*}
\endgroup

Constraining gas motions across clusters in different dynamical states is important for understanding the role of turbulence and bulk motions in the evolution of the ICM. In cluster cores, these motions may contribute to offsetting radiative cooling through turbulent dissipation or mixing \citep[e.g.,][]{Zhuravleva2014,XRISM_Perseus2026,M87_paper1}, while at larger radii gas motions contribute to nonthermal pressure support and departures from hydrostatic equilibrium \citep[e.g.,][]{Lau2009,Nelson2014b}. Comparisons between relaxed and merging systems, and AGN-active vs. inactive systems, therefore provide insight into the relative roles of AGN feedback and mergers in regulating the state of galaxy clusters. 

Abell 3571 (A3571) is a massive \cite[$M_{\mathrm{tot}} = 4.6\times 10^{14}\ h^{-1}\ M_{\odot}$,][]{Nevalainen2000} member of the Shapley supercluster \citep{Raychaudhury1991} and the sixth brightest galaxy cluster in the X-ray sky \citep{Edge1990}. Its large central cD galaxy, MCG\,05-33-002 \citep{KempMeaburn1991}, at $z=0.0386$ \citep{Hudson2001}, is elongated in the north-south direction, matching the morphology of the cluster as a whole \citep{Quintana1993} and the alignment of the cosmic filaments in the region \citep{Raychaudhury1991}. Although the elongation of the cluster may be taken as an indicator that it is not relaxed, X-ray studies of its temperature and metallicity distributions did not find prominent deviations from azimuthal symmetry \citep{Markevitch1998}. Multiple works suggest that A3571 is in a late- or post-merger state rather than undergoing an active major merger \citep{Venturi2002, Rossetti2010, Zheng2026}. A central cool component was reported in ASCA observations, but later XMM-Newton studies identified areas of uneven temperature rather than a classic cool-core temperature drop \citep{Markevitch1998, Hudaverdi2005}. As a result, A3571 has been classified as a weak cool-core cluster, intermediate between a cool core and non-cool core cluster \citep{Rossetti2010, Frank2013}. Despite its relatively short central cooling time ($7.7$\,Gyr within 55 kpc \citep{Birzan2012}), there is no unambiguous evidence for AGN-inflated cavities in Chandra images \citep{Birzan2012}, although \cite{Olivares2023} reported potential cavities, and \cite{Venturi2002} identified a faint 22 cm radio source associated with MCG\,05-33-002.

A3571, therefore, provides an opportunity to investigate ICM turbulence in a relatively relaxed system where merger-induced motions, manifested primarily as gas sloshing, may dominate over AGN feedback. The cluster was observed with four XRISM/Resolve pointings during general observer (GO) cycle 1, which forms the basis of this work.

This paper is organized as follows. In Section 2, we introduce the XRISM observations and describe the data reduction steps. Section 3 presents the data analysis methods, including the spatial binning strategies and spectral models considered. In Section 4, we highlight the kinematic measurements and derived gas characteristics, together with details of the tests of potential systematic uncertainties. Section 5 discusses the implications of the measurements, including constraints from resonant scattering, comparisons with other clusters observed by XRISM, estimates of cooling and heating rates, and comparisons with cosmological simulations. Section 6 contains a summary of the work and concluding remarks.

Throughout this paper, we assume the Planck 2018 cosmology \citep{Planck2020}. At the average cluster redshift of 0.039 \citep{Mahdavi2004}, this corresponds to a scale of 0.8 kpc/$1\arcsec$.

\begin{center}
    \begin{figure*}
        \centering
    	\includegraphics[width=0.95\linewidth]{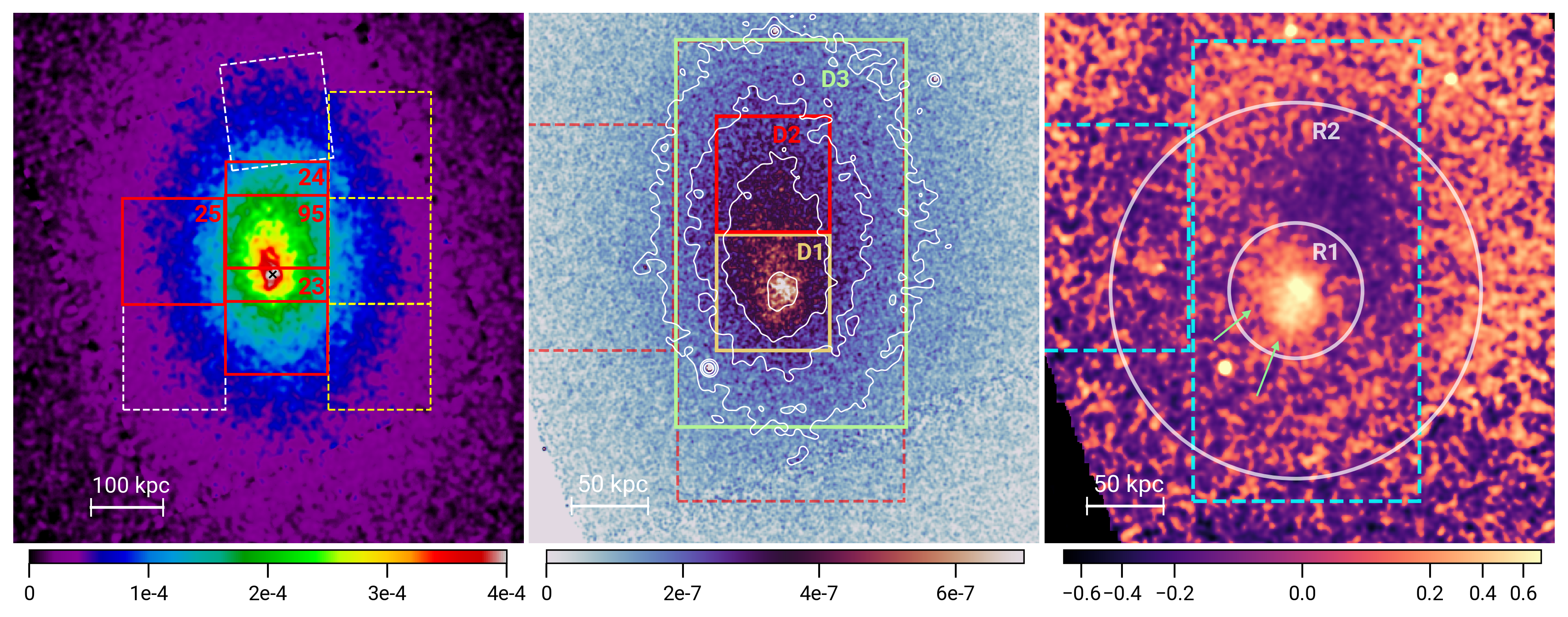}
    	\caption{Left: 2-8 keV XMM-Newton image with 2\arcsec\ binning used for ARF creation, displayed with a colormap that emphasizes the X-ray surface brightness contours. Red solid boxes are cycle 1 observations used in this analysis, labeled in their upper right corners. Additional observations observed in cycle 2 (white dashed) and accepted in cycle 3 (yellow dashed, roll angle unknown) will be the subject of an upcoming paper. The location of the BCG is marked with an `X'. Center: Gaussian-smoothed ($\sigma=1$ pixel or 0.983$\arcsec$) 0.5-8 keV Chandra image of A3571 with contours overlaid. The pixels used in the default binning regions (see Sec.~\ref{sec:bin}) are identified with yellow, red, and green respectively. The area covered by all cycle 1 pointings is outlined with red dashed lines. Right: Heavily-smoothed 0.5-8 keV Chandra residual image (image divided by the best-fit double elliptical $\beta$ model) altered to emphasize the faint spiral structure. No AGN structures are visible; a cold front in R1, to the southeast of the BCG, is labeled with green arrows. The ``sky'' regions corresponding to the second (radial) binning strategy are shown in white and the area covered by all cycle 1 pointings is outlined with cyan dashed lines.}
        \label{fig:image} 
    \end{figure*}
\end{center}

\section{Observations and Data Reduction}
\label{sec:obs}

Abell 3571 (A3571) was observed in four XRISM pointings during GO cycle 1 between December 30, 2024 and January 9, 2025. The Resolve observations were conducted with an open filter wheel and closed gate valve. The pointing configuration and exposure time of each observation are listed in Table~\ref{tab:observations} and the pointings are shown overlaid on an XMM-Newton image in Figure~\ref{fig:image}, left panel. The creation of this image is described in Appendix~\ref{subsec:models}. The overlaid observations used in this work are shown in red, whereas additional cycle 2 observations of A3571 are shown in white, and accepted cycle 3 observations are shown in yellow (roll angle as yet unknown). This work focuses on the core of A3571; a concurrent paper from our team focuses on large-scale sloshing motions via velocity mapping in the cycle 1 (red) observations \citepalias{Aihara2026}. All of the XRISM observations, including cycle 2 and 3 observations, will be discussed in an upcoming paper.

The Resolve data reduction was performed with HEASoft v.~6.35.1 and calibrated with Resolve CalDB v.~11 (20250315). The task \texttt{xapipeline} was first run on each observation to reprocess the data and produce a clean event file. The standard proximity and rise-time screenings were then applied, following the XRISM Quick-Start Guide v.~2.3 \footnote{\url{https://heasarc.gsfc.nasa.gov/docs/xrism/analysis/quickstart/index.html}}. Images, light curves, and spectra containing only high resolution primary (Hp) events were then extracted for each subset of pixels using \texttt{xselect}. Due to anomalous energy-scale jumps during the observations, pixel 27 was excluded from ObsIDs 201023010, 201024010, and 201025010, and pixel 29 was excluded from 201024010 and 201095010. Pixel 12 is a calibration pixel and is therefore excluded in all cases. The light curves were checked for potential solar flare contamination and were found to have none (for more details, see Appendix~\ref{ap:flare}). 

The next step was to create the spectral response files. We first created the redistribution matrix files (RMFs) using \texttt{rslmkrmf}. We used as input a cleaned event file whose low-resolution secondary (Ls) events had all been removed. Large RMFs, which account for the entire line spread function, were extracted. An attitude histogram and exposure map were created with \texttt{xaexpmap}. Auxiliary response files (ARFs) were calculated with the \texttt{xaarfgen} tool in IMAGE mode, using an exposure-corrected 2.0-8.0 keV XMM-Newton image with $2 \arcsec$ binning as the default input (Appendix~\ref{subsec:models}). It is crucial to model the spatial-spectral mixing (SSM), so ARFs were made not only for each subset of pixels but also to account for the cross-scattering of photons between regions. The definition of regions used for spectral extraction is discussed in Section~\ref{sec:analysis}.

The non-X-ray background (NXB) spectra were estimated from the second version of Resolve's night-Earth database using the \texttt{rslnxbgen} task. Filters were applied to the database to match the pixels, cut-off rigidity distribution, and event screening criteria for each A3571 region. 

The spectra for each region, including the NXB spectra, were optimally binned with \texttt{ftgrouppha} following \cite{Kaastra2016}.

\section{Data Analysis}
\label{sec:analysis}

\subsection{Spatial binning strategies}\label{sec:bin}
To choose regions for analysis, we first used the tool \texttt{xmatraceback} to estimate the amount of scattered light between different groups of pixels. Our aim was to use the smallest possible regions while maintaining the ability to model the cross-scatter, meaning that the region size was limited (i) to the approximate size of Resolve's Half Power Diameter \citep[1.3\arcmin;][]{Tashiro2025}, (ii) to cases where scattered light did not dominate the measured photons in a region (i.e. was $<50\%$ of total measured photons), and (iii) by the necessity of having enough photons for velocity broadening measurements. For our goal of comparing the core of A3571 to feedback-active cores, we chose a region centered on the brightest cluster region (so-called Region D1); a region to the north, following the surface brightness contours (Region D2); and a region radially outside of these regions, also following the surface brightness contours (Region D3). The central panel of Figure~\ref{fig:image} shows the 0.5-8 keV Chandra image (whose creation is described in Appendix~\ref{subsec:models}) covering the core of A3571 with surface brightness contours and the default binning strategy overlaid. Sky areas beyond these three regions can be safely excluded from the SSM analysis, as they contribute $<10\%$ of each region's photons. In each case, the ``sky region'' used as image input for ARF ray-tracing was taken to be the same as the area covered by the ``detector region'', i.e. the area covered by the pixels. 

Although the default binning choice represents the finest binning that still fulfills recommendations to reduce cross-scattering, other binning approaches have unique advantages. To compare the radial profile of A3571 with other clusters, it was useful to adopt a binning strategy whose regions are symmetric around the cluster's core.  Cross scattering requirements allowed for two regions (R1 and R2) whose sky regions (image used for ARF creation) are a circle of radius $0-0.9\arcmin$ and an annulus spanning $0.9-2.5\arcmin$ respectively. The sky regions are shown overlaid on a 0.5-8 keV Chandra residual image (image divided by best-fit elliptical $\beta$ model) in the right panel of Figure~\ref{fig:image}. Sky regions beyond these had a negligible contribution to total photon count rate, and could therefore be ignored.

The ObsIDs and pixels used in each binning strategy are listed in Table~\ref{tab:regions}.

\begingroup 
    \setlength{\tabcolsep}{6pt}
    \renewcommand{\arraystretch}{1} 
    \setlength\extrarowheight{2pt}
    \begin{table}
        \centering
        \begin{tabularx}{\linewidth}{ c l l l }  
            Config. & Reg. & ObsID & Detector pixels \\
            \hline \hline
            Default & D1 & {'95} & 9-10, 17-22, 25\\
             & & {'23} & 1-2, 28-29, 31\\
             & & {'24} & 9, 19, 21 \\
             & D2 & {'95} & 0-2, 27-28, 31, 33, 35 \\
             & & {'24} & 0, 10, 17-18, 20, 22, 25, 33, 35 \\
             & D3 & {'95} & 3-8, 11, 13-16, 23-24, 26, 30, 32, 34\\
             & & {'23} & 0, 3-8, 15-18, 25-26, 30, 32-35 \\
             & & {'24} & \makecell[tl]{1-8, 11, 13-16, 23-24, 26, 28, 30-32,\\ 34} \\
            Radial & R1 & {'95} & 9-10, 17-22, 25 \\
            & & {'23} & 1-2, 28-29, 31 \\
            & & {'24} & 9, 19, 21 \\
             & R2 & {'95} & 0-8, 11, 13-16, 23-24, 26-28, 30-35\\
             & & {'23} & \makecell[tl]{3-8, 10, 13, 15-18, 20, 22, 24-26, 30,\\ 32-35} \\
             & & {'24} & 0, 7, 10-11, 13-18, 20, 22-26, 33-35 \\
             & & {'25} & 6-8, 11, 13-16 \\
             \hline \hline  
        \end{tabularx}  
        \caption{Details of the region divisions used for spectral analysis.}
        \label{tab:regions}
    \end{table}
\endgroup

\subsection{Spectral modeling}
\label{subsec:spec}

We fit the Resolve spectra using Xspec v. 12.15.0 \citep{Arnaud1996} and atomic database AtomDB 3.1.3. All abundances are relative to \cite{Lodders2009} proto-solar abundances. Best-fit values were determined by minimizing the C-statistic \citep{Cash1979,Wachter1979}. 

The first step was to model the NXB spectra for each region. This was done using a diagonal RMF and a power-law model with Gaussians for known detector emission lines{\interfootnotelinepenalty10000\footnote{\url{https://heasarc.gsfc.nasa.gov/docs/xrism/analysis/nxb/nxb_spectral_models.html}}}.  The fitting was performed in the 2-11 keV band, by first fitting the overall normalization of the spectrum and then fitting the normalizations of the individual lines. The best-fit NXB models were then fixed and included in the spectral fits of the source.

\begin{figure*}
    \centering
    \includegraphics[width=0.9\linewidth]{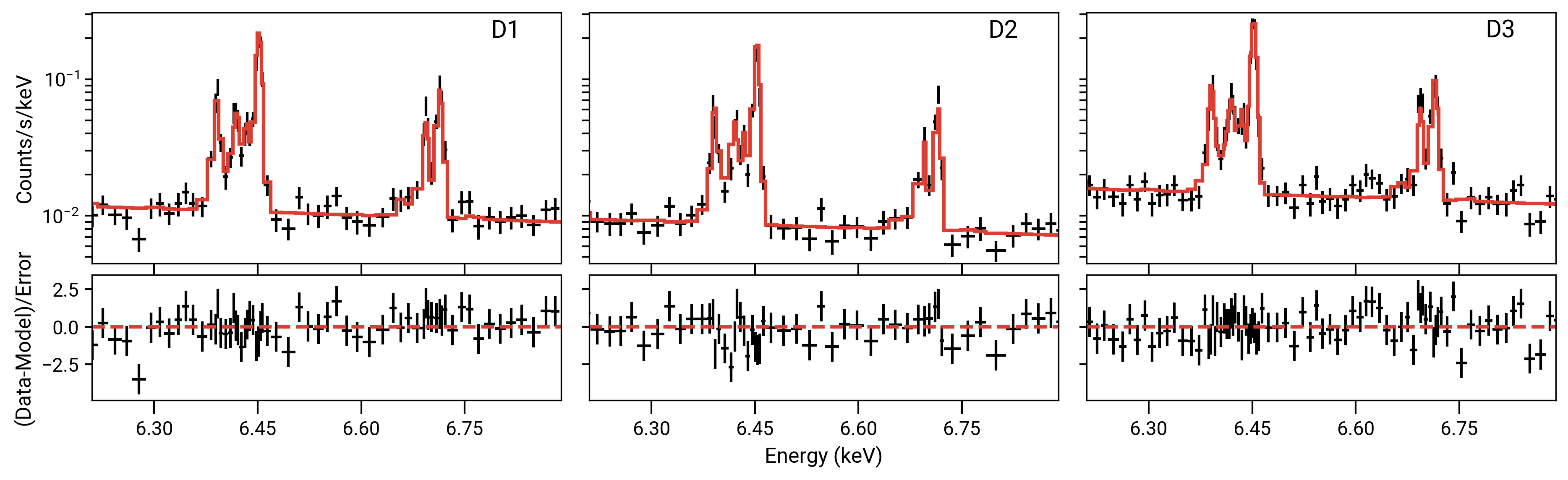}
    \caption{Spectra, best-fit models, and residuals for the default-binning regions D1, D2, and D3. Fits were performed in the 2-11 keV band, but spectra are shown in a narrower band where there are prominent Fe emission lines. Spectra were binned to $>5\sigma$ significance for visual clarity.}
    \label{fig:spectra}
\end{figure*}

For the source spectra, an energy band of 2-11 keV was used. A single-temperature, collisionally-ionized thermal plasma model affected by Galactic absorption (in Xspec, \texttt{TBabs*bvapec}) was used to model the ICM in each region. All regions were fit simultaneously, accounting for the SSM. The Galactic hydrogen column density was fixed to $3.88\times10^{20}$ cm$^{-2}$ \citep{HI4PI2016}. The temperature, redshift ($z$), velocity dispersion ($\sigma_v$), normalization, and abundance of Fe were treated as free parameters. The abundances of He and C were fixed to 1.0 Solar, and all other elements were tied to Fe abundance. Figure~\ref{fig:spectra} shows the spectra around the Fe K and Ly$\alpha$ lines and best-fit models for the default binning. The full 2-11\,keV band showing the contributions due to SSM from various regions can be found in Figure~\ref{fig:broadband}.

In addition to the baseline model, we also considered the case where other elements were allowed to vary. Although less significant than Fe, lines associated with Ni, Ar, and Ca are also visible, to varying significance, in the A3571 spectra (Fig.~\ref{fig:broadband}). Additionally, the data were fitted with a combination of two thermal plasma models (\texttt{TBabs(bvapec+bvapec)}). Depending on the desired number of free parameters, all parameters of the second \texttt{bvapec} model are tied to the first except for a) temperature; b) temperature and velocity dispersion; and c) temperature, velocity dispersion, and Fe abundance.

We also accounted for the case that resonant scattering may be present. Resonant scattering would mainly affect the shape and flux of the strongest W line in D1, corresponding to the central region. We therefore removed the W line from the atomic database and replaced it with a Gaussian in the model prior to fitting.

\section{Results}
\label{sec:results}

\begin{figure*}
    \centering
    \includegraphics[width=0.47\linewidth]{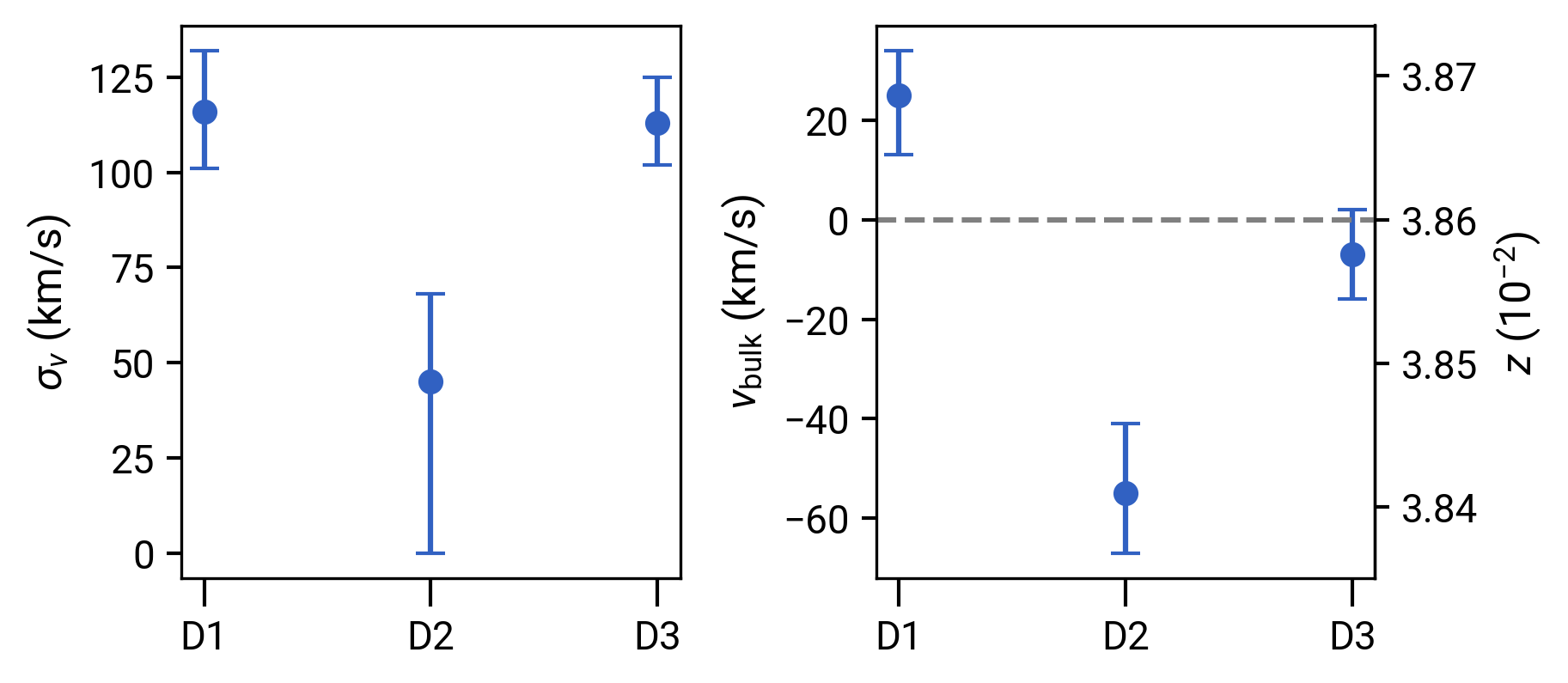}
    \includegraphics[width=0.45\linewidth]{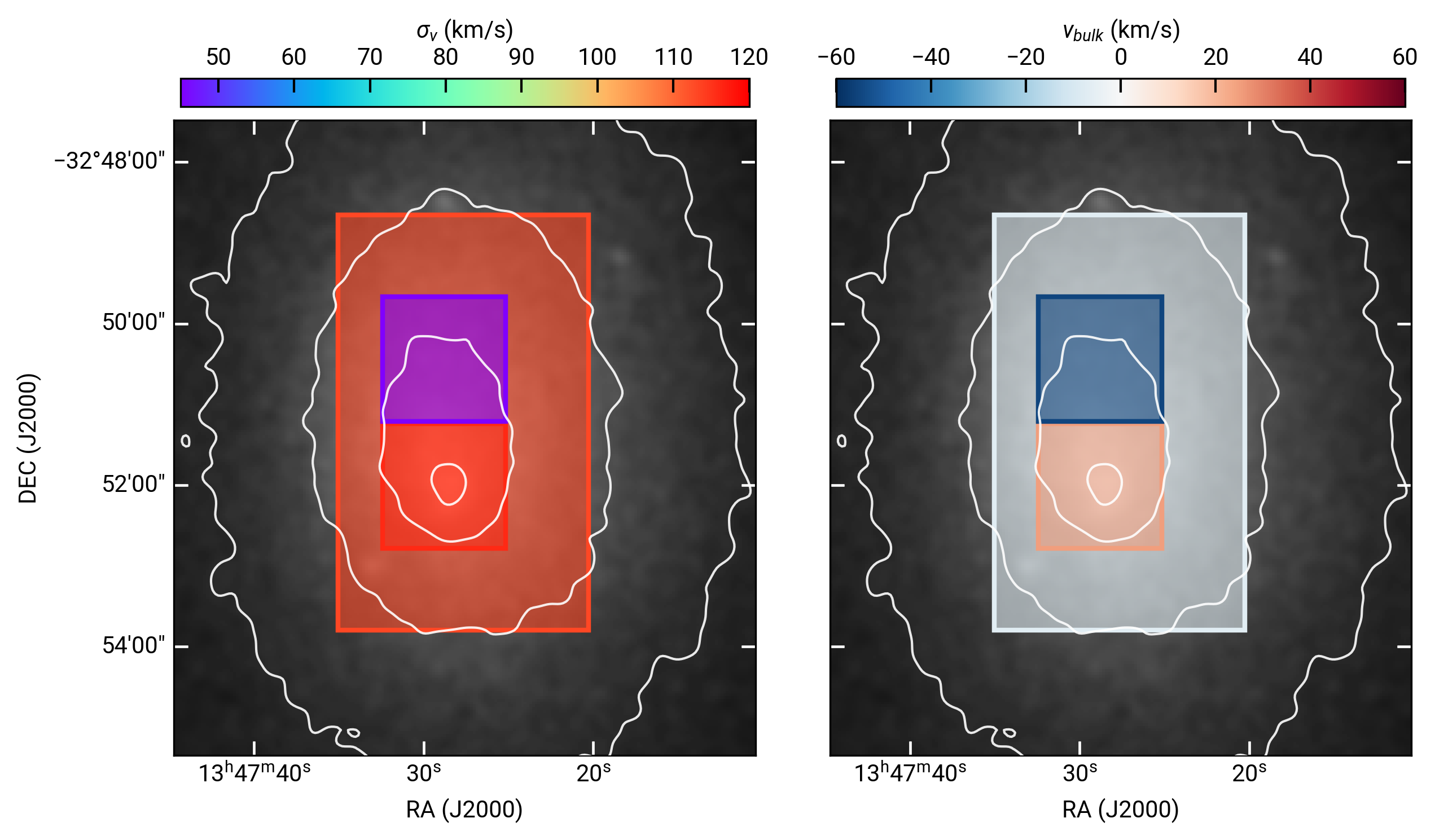}
    \includegraphics[width=0.47\linewidth]{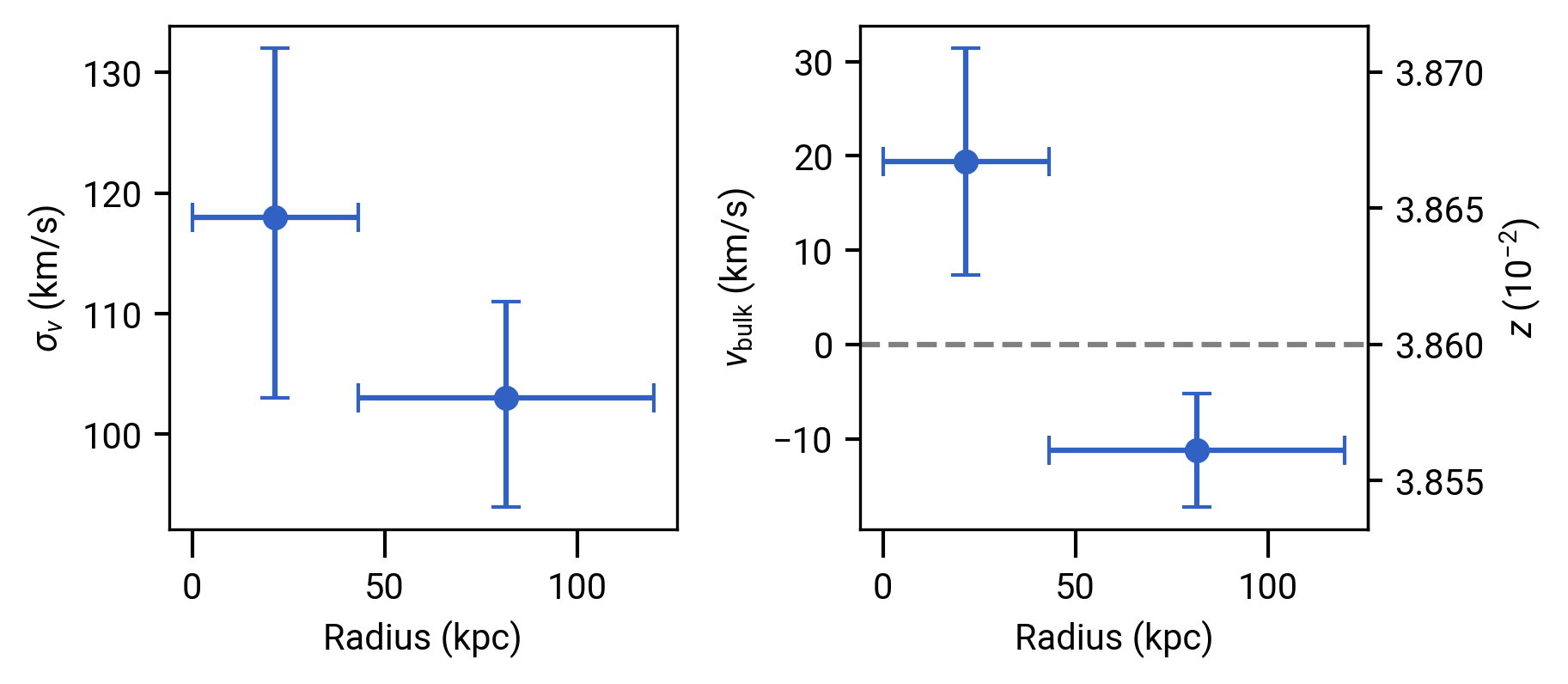}
    \includegraphics[width=0.45\linewidth]{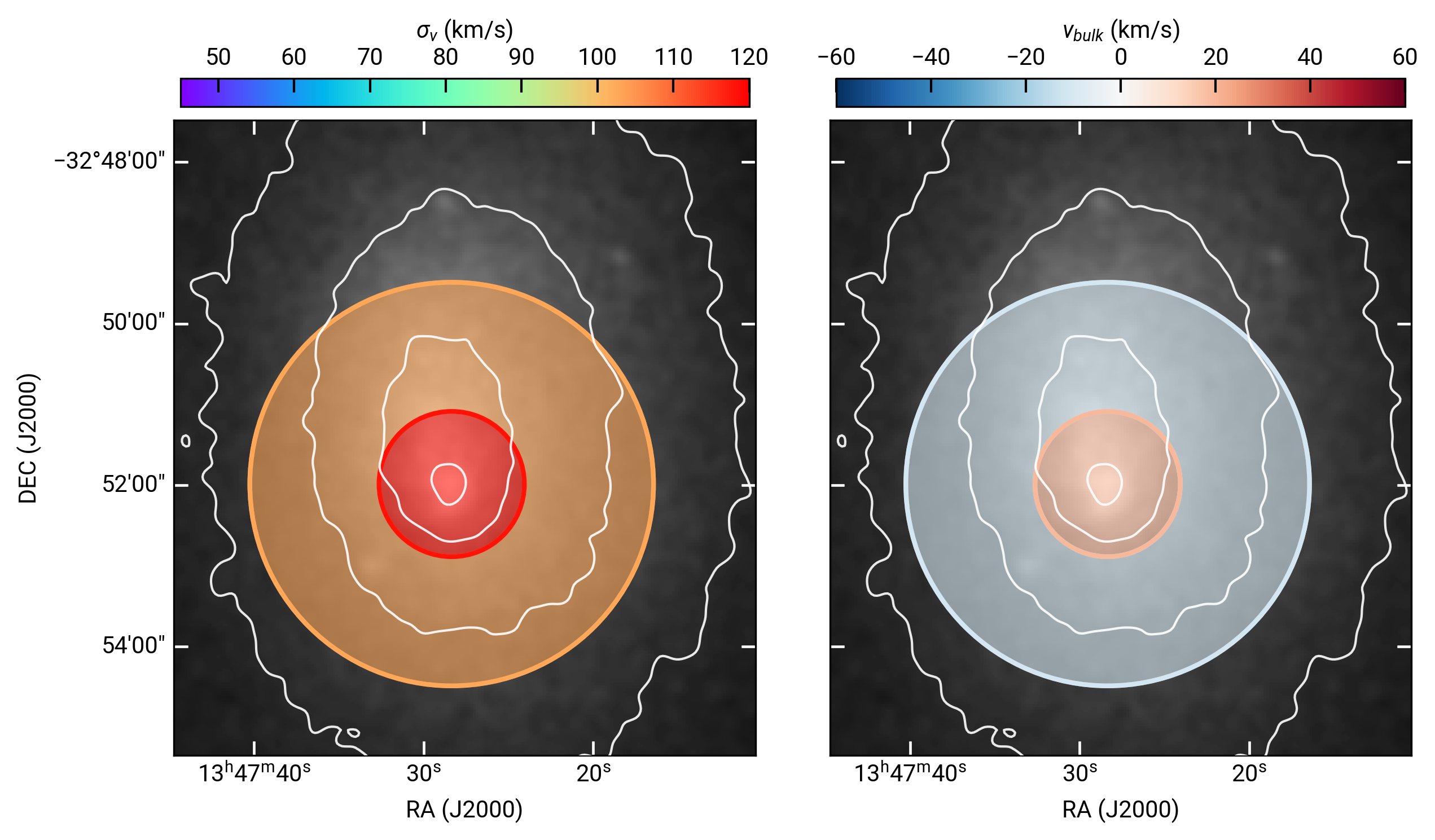}
    \caption{Left: velocity dispersion and heliocentric-corrected bulk velocity/ redshift with respect to the BCG as a function of radius for (top) the default region binning and (bottom) the radial region binning. Right: Same quantities but now shown in spatial regions with surface brightness contours.}
    \label{fig:params}
\end{figure*}

The top row of Figure~\ref{fig:params} shows the best-fit parameters for the default binning, for which the C-statistic per degree of freedom (C-stat/d.o.f.) is 16068/16370. The $\sigma_v$ in D1, covering the core of the cluster, is $\sim116~\mathrm{km~s^{-1}}$. For the region to the north, D2, which covers the elliptical extension of the cluster, we obtain $\sigma_v=45^{+23}_{-45} ~\mathrm{km~s^{-1}}$, consistent with zero at the 1$\sigma$ level. The region radially outside of these two regions, D3, shows a similar $\sigma_v$ to D1 at $\sim 113~\mathrm{km~s^{-1}}$.

The best-fit redshifts for regions D1-D3 are $3.86\times10^{-2}$, $3.83\times10^{-2}$, and $3.85\times10^{-2}$. Bulk velocity can be calculated from the redshift using $v_{\mathrm{bulk}} = c(z - z_{BCG})/(1 + z_{BCG})$ with the redshift of the BCG $z_{\textrm{BCG}}=0.0386$ \citep{Hudson2001}. Heliocentric corrections are found with the \texttt{barycen} HEASoft tool and subtracted from the bulk velocity. The mean correction is $25.2\,\mathrm{km~s^{-1}}$ for '95, 27.1\,$\mathrm{km~s^{-1}}$ for '23, 26.3\,$\mathrm{km~s^{-1}}$ for '24, and 26.3$\,\mathrm{km~s^{-1}}$ for '25.  The bulk velocity of the system with respect to the BCG following the correction is then 25\,$~\mathrm{km~s^{-1}}$ in D1, -55\,$~\mathrm{km~s^{-1}}$ in D2, and -7\,$~\mathrm{km~s^{-1}}$ in D3, with the heliocentric-corrected redshifts $3.87\times10^{-2}$ (D1), $3.86\times10^{-2}$ (D2), and $3.84\times10^{-2}$ (D3). In the upper right panel of Figure~\ref{fig:params}, D1 and D3 exhibit similar velocities near 0\,$~\mathrm{km~s^{-1}}$, while D2 is blueshifted. The trend in $v_{\mathrm{bulk}}$ mimics that in $\sigma_v$.

The bottom row of Figure~\ref{fig:params} shows the best-fit parameters for the radial binning strategy. As one might expect using the same pixels, core region R1 displays a velocity dispersion comparable to D1 of $\sim 118$\,$~\mathrm{km~s^{-1}}$. Region R2 displays only a slight decrement to $\sim 103$\,$~\mathrm{km~s^{-1}}$, which falls within 1$\sigma$ of the R1 measurement. The bulk velocity/redshift behave likewise; R1 is slightly redshifted at $\sim 19$\,$~\mathrm{km~s^{-1}}$ with respect to the BCG, while R2 is slightly blueshifted, $\sim -11$\,$~\mathrm{km~s^{-1}}$. All best-fit values with uncertainties for the default and radial binnings fit with the baseline spectral model are presented in Table~\ref{tab:results}.

In the case of the more complex models that added temperature, abundance, and/or velocity components, no improvement in the fit statistic (C-stat/d.o.f.) was obtained relative to the baseline model. The two-temperature model yielded temperatures of 4.42$\substack{+0.85 \\ -0.47}$\,keV and 9.04$\substack{+0.56 \\ -0.91}$\,keV, while maintaining the same velocity values as the baseline fit. The best-fit temperature values were stable regardless of whether one or two velocity components were included.

Allowing elements besides Fe to vary did not provide a statistically improved fit, although Ar, Ca, and Ni were detected at $\geq2\sigma$ significance in D1/R1. The best-fit values are $1.19\pm0.41$ for Ar, $1.12\pm0.33$ for Ca, and $0.63\pm0.23$ for Ni. Elements significantly detected in outer regions were Ni ($0.52\pm0.10$ in R2, $0.59\pm0.22$ in D2, $0.53\pm0.13$ in D3) and Ca ($0.41\pm0.14$ in R2). Leaving these abundances free to vary did not affect the kinematic parameters.

Using the resonant scattering model did not lead to an improvement in the test statistic. The best-fit kinematic parameters were unchanged within $1\sigma$. More thorough resonant scattering model results are presented in Section~\ref{subsec:rs}.

\subsection{Derived ICM properties}

From the measured ICM temperature and velocities, several gas characteristics can be calculated. The sound speed is given by $c_s = \sqrt{\gamma k_\mathrm{B} T/(\mu m_p)}$, where $\gamma = 5/3$ is the adiabatic index for monatomic gas, $k_B$ is the Boltzmann constant, $T$ is the measured gas temperature, $\mu=0.61$ is the mean molecular weight, and $m_p$ is the proton mass. In the case of radial binning for A3571, $c_s$ is $1316\pm 19$\,$~\mathrm{km~s^{-1}}$ and $1315\pm10~\mathrm{km~s^{-1}}$ for R1 and R2, respectively, showing no radial trend.

The sound speed can then be used to find the 3D Mach number, $M_\mathrm{{3D}}=\sqrt{3}\sigma_v/c_{\rm s}$, under the assumption that the line-of-sight velocity dispersion $\sigma_v$ is representative of an isotropic turbulent velocity field. The non-thermal (kinetic) pressure fraction is then $P_{\rm NT}/P_{\rm tot}=M_{\mathrm{3D}}^2/({M_{\mathrm{3D}}^2 + 3/\gamma})$.

\begin{figure}
    \centering
    \includegraphics[width=\linewidth]{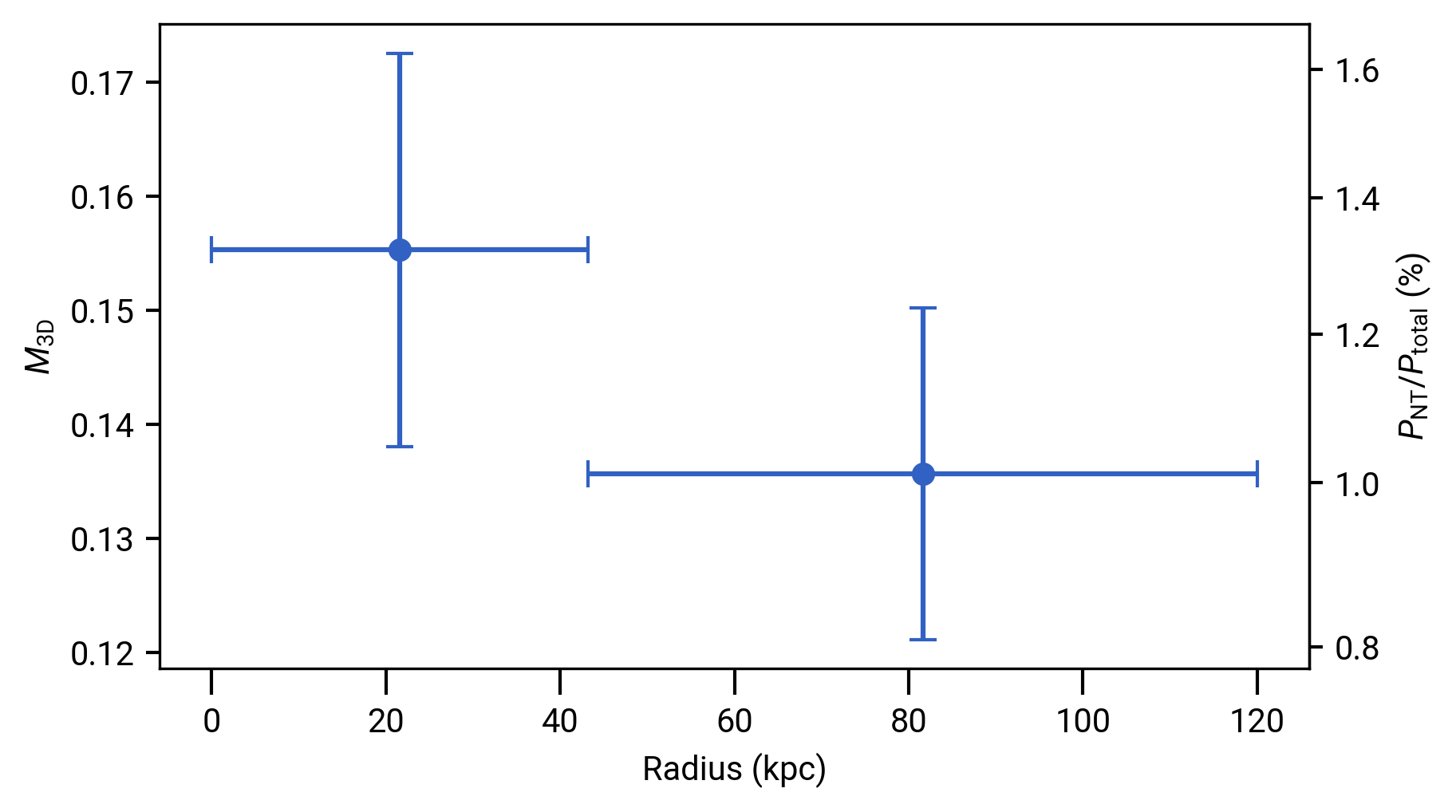}
    \caption{3D Mach number $M_{\mathrm{3D}}$ (left Y-axis) and non-thermal pressure fraction $P_{\rm NT}/P_{\rm tot}$ (right Y-axis) for the radial binning strategy, shown with $1\sigma$ uncertainties. No strong radial trends are apparent.}
    \label{fig:derived}
\end{figure}

Figure~\ref{fig:derived} shows $M_{\mathrm{3D}}$ on the left y-axis and $P_{\rm NT}/P_{\rm tot}$ on the right y-axis for the radial binning strategy. We find $M_{\mathrm{3D}}=0.16\pm0.02$ and $P_{\rm NT}/P_{\rm tot}=1.3\pm0.3\%$ for R1, and $M_{\mathrm{3D}}=0.14\pm0.01$ and $P_{\rm NT}/P_{\rm tot}=1.0\pm0.2\%$ for R2. We do not detect a radial trend in $M_{\mathrm{3D}}$ or $P_{\rm NT}/P_{\rm tot}$; the values of the two regions agree within $1\sigma$.

\subsection{Systematics checks}

To verify the robustness of our reported results, we also explored the effects of 
\begin{itemize}
    \item \textit{Choice of band:} We performed fits in the 3-11 keV and 4-7 keV bands, and found no significant change in the best-fit kinematic parameters compared to the baseline 2-11 keV band.
    \item \textit{AGN presence:} Some previous literature characterizes A3571 as having a central AGN due to its faint radio emission \citep{Frank2013}, but no point source is visible in the hard X-ray band image from Chandra. To cover the possibility that an AGN may contribute to the spectrum, a power law was included in addition to the baseline model in D1. Its photon index was constrained to the range 1.5--2.5 and its normalization was free. This did not result in an improvement in the test statistic.
    \item \textit{SSM:} Using the baseline model but excluding the cross-ARFs that account for the SSM, we find $\sigma_v$ and $z$ values that differ by $\sim 2 \sigma$ in the D1 and D2 regions. However, the best-fit values are of the expected level given the amount of estimated cross-scatter between the regions, indicating the reliability of the SSM correction.
    \item \textit{Atomic data table:} Using SPEXACT atomic data files v 3.08.01 (via option \texttt{APECUSESPEX} in Xspec), we find no significant effect on the best-fit values as compared to the default AtomDB tables.
    \item \textit{RMF choice:} Using XL sized RMFs, which in addition to the components included in the size-L RMFs account for electron loss continuum, does not have an effect on the best-fit values.
    \item \textit{ARF input image:} In addition to the XMM images, 2-8 keV Chandra sky images were used as input to the raytracing for ARF creation. This resulted in no change to the kinematic parameters.
    \item \textit{Binning strategy:} Other binning strategies were tested to check the effect of region selection on the best-fit parameters. In particular, following the same binning strategy as \cite{Aihara2026} in the core region, we found comparable values within $1\sigma$. Their core region of four pixels (their ``region 5'') displays a lower $\sigma_v$ than our core region of nine pixels ($\leq60~\mathrm{km~s^{-1}}$ vs $115\pm15~\mathrm{km~s^{-1}}$). This difference arises from the different pixel selections used to define the ``central'' region: ``region 5'' is offset from the X-ray peak by approximately $19\arcsec$ (15 kpc), whereas we selected our radial regions to place the X-ray peak closer to their center. This choice is better suited to our goal of studying the core of the cluster. Section~\ref{subsec:heat} includes a discussion on how using the four-pixel measurement affects calculated heating rates and resulting conclusions.
\end{itemize}

\begingroup
    \setlength{\tabcolsep}{10pt}
    \renewcommand{\arraystretch}{1.5}
    \setlength\extrarowheight{2pt}
    \begin{table*}
        \centering
        \begin{tabular}{ c c c c c c c c}  
            Region & kT (keV) & $\sigma_v$ ($\mathrm{km~s^{-1}}$) & $z$ ($10^{-2}$) & $v_{\mathrm{bulk}}$ ($\mathrm{km~s^{-1}}$) & Fe & $M_{\mathrm{3D}}$ & $P_{\rm NT}/P_{\rm tot} (\%)$\\
            \hline \hline     
            D1 & 6.60 $\substack{+0.18 \\ -0.19}$ & 116 $\substack{+16 \\ -15}$ & 3.869 $\substack{+0.003 \\ -0.004}$ & 25 $\substack{+9 \\ -12}$ & 0.61 $\substack{+0.04 \\ -0.04}$ & $0.15\substack{+0.02 \\ -0.02}$ & $1.3\substack{+0.4 \\ -0.3}$ \\
            D2 & 5.86 $\substack{+0.18 \\ -0.17}$ & 45 $\substack{+23 \\ -45}$ & 3.841 $\substack{+0.005 \\ -0.004}$ & -55$\substack{+14 \\ -12}$ & 0.50 $\substack{+0.04 \\ -0.04}$ & $0.06\substack{+0.03 \\ -0.06}$ & $0.2\substack{+0.2 \\ -0.2}$ \\
            D3 & 6.81 $\substack{+0.12 \\ -0.12}$ & 113 $\substack{+12 \\ -11}$ & 3.858 $\substack{+0.003 \\ -0.003}$ & -7$\substack{+9 \\ -9}$ & 0.49 $\substack{+0.02 \\ -0.02}$ & $0.15\substack{+0.01 \\ -0.02}$ & $1.2\substack{+0.2 \\ -0.2}$ \\
            R1 (0-0.9$\arcmin$) & 6.62$\substack{+0.19 \\ -0.19}$ & 118$\substack{+14 \\ -15}$& 3.867$\substack{+0.004 \\ -0.004}$& 19.4 $\substack{+12 \\ -12}$ & 0.61$\substack{+0.04 \\ -0.03}$ &  $0.16\substack{+0.02 \\ -0.02}$  & $1.3\substack{+0.3 \\ -0.3}$\\
            R2 (0.9-2.5$\arcmin$) & 6.61$\substack{+0.09 \\ -0.09}$ & 103$\substack{+8 \\ -9}$ & 3.856$\substack{+0.002 \\ -0.002}$ & -11.2$\substack{+6 \\ -6}$ & 0.49$\substack{+0.01 \\ -0.01}$ & $0.14\substack{+0.01 \\ -0.01}$ & $1.0\substack{+0.2 \\ -0.2}$\\
             \hline \hline  
        \end{tabular}  
        \caption{Best-fit values and 1$\sigma$ errors for chosen configurations. Redshifts and bulk velocities include heliocentric correction.}
        \label{tab:results}
    \end{table*}
\endgroup

\section{Discussion}
In addition to velocity broadening measurements, resonant scattering techniques can also provide a measure of turbulent motions in the ICM. In this section, we present our application of resonant scattering simulations to the data. We then compare our findings with other clusters that have been observed with XRISM, and with predictions from cosmological simulations. In the final subsection, we calculate the turbulent heating rate for comparison with the radiative cooling rate, and discuss the implications.

\subsection{Resonant scattering}
\label{subsec:rs}

Some of the strongest emission lines in the spectra of galaxy clusters have non-negligible optical depths, leading to resonant scattering \citep[e.g.,][]{Gilfanov1987,1989ApJ...345...12S,Sazonov2002,Churazov2010,Ogorzalek2017,HitomiRS2018, Tanaka2026}. Gas motions in the ICM broaden these lines, thereby reducing the effective optical depth and suppressing line distortions produced by scattering. Thus, resonant scattering is used as a diagnostic of gas motions in the ICM independent of Doppler broadening measurements. Previous studies have also shown that resonant scattering is primarily sensitive to small-scale random motions in the gas \citep{Zhuravleva2011}.

Using deprojected radial profiles of gas density, temperature, and heavy element abundances in A3571 (discussed in Appendix~\ref{subsec:models}), we estimate the radially-integrated optical depth at the line center of the resonant Fe XXV $w$-line, one of the most promising transitions for resonant scattering at these temperatures, to be $\sim 0.53 \ (0.34)$ for $M_{\mathrm{3D}}=0 \ (0.15)$. The non-zero Mach number was selected based on the line-broadening measurements in the core. The optical depth is sufficiently large for resonant scattering to suppress the $w$-line flux in the central region of A3571, even in the case of $M_{\mathrm{3D}}=0.15$.

\begin{figure}
    \centering
    \includegraphics[width=0.95\linewidth]{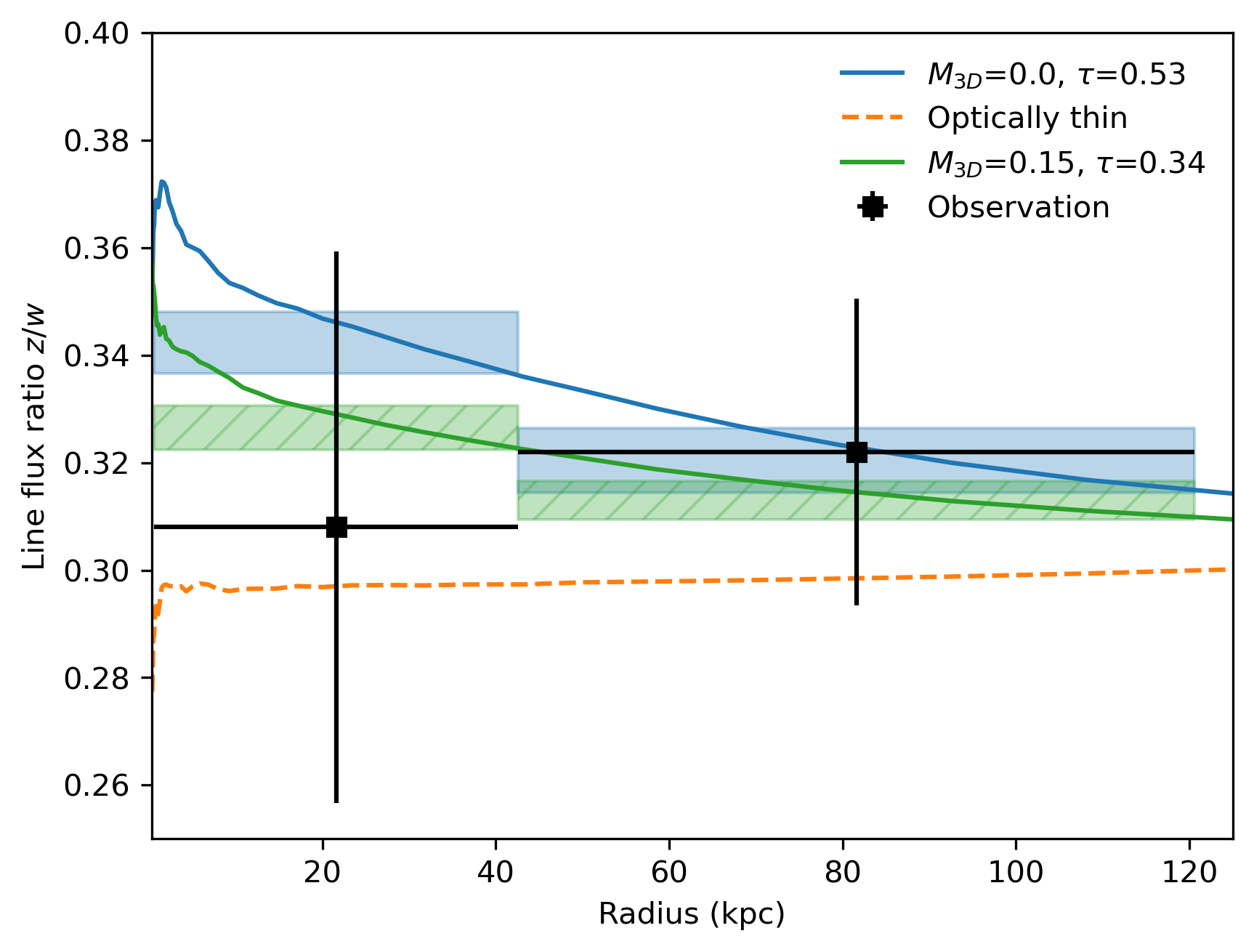}
    \caption{The ratio of line fluxes $z/w$ as a function of radius. The blue (green) curve shows the case for $M_{\mathrm{3D}}=0(0.15)$ with resonant scattering, while the orange curve shows the optically thin case. The shaded regions show the average, emissivity-weighted flux for each bin. The black points represent the flux ratios from the best-fit model to XRISM observations.}
    \label{fig:rs}
\end{figure}

We then performed a series of tailored Monte Carlo simulations of resonant scattering, assuming a spherically symmetric cluster model and isotropic small-scale motions characterized by a given Mach number. Details of the simulations are described in several works, including \cite{Sazonov2002,Zhuravleva2013, HitomiRS2018}.  Figure~\ref{fig:rs} shows that these simulations predict an enhancement of the ratio of fluxes in forbidden-to-resonant lines, $z/w$, from the optically thin value of $\sim 0.3$ (dashed curve) to $\sim 0.37$ in the static-gas case (blue curve). As the characteristic Mach number of the gas increases (green curve), the predicted line ratio gradually approaches the optically thin limit. 

We measure the $z/w$ flux ratio in A3571 by modeling the Fe XXV $w$ and $z$ lines with Gaussian components within the plasma emission model, obtaining values of $0.31\pm 0.05$ and $0.32\pm 0.03$ in the central and adjacent radial regions, respectively. These measurements are indicated by the black points in Figure~\ref{fig:rs}. To compare directly with the observations, we compute an emissivity-weighted average of the simulated curves over the same spatial regions used in the observations (shaded regions). 
In the central region (within $\sim 40$\,kpc), changing the Mach number from $0$ to $0.15$ results in a decrease in the $z/w$ ratio from $\sim0.34$ to $\sim0.33$. Increasing the Mach number further would bring the ratio closer to the optically thin case with $z/w \sim 0.3$. With increased photon statistics, XRISM data could distinguish these cases. One can see that the current $z/w$ ratios are consistent with a broad range of Mach numbers within the current statistical uncertainties, although the central value is closer to the optically thin limit (higher Mach number).

A3571, therefore, appears to be a promising target for resonant scattering studies, complementing detections in the Perseus \citep{XRISM_Perseus2026} and PKS 0745-191 \citep{Tanaka2026} clusters. Future observations of A3571 with improved photon statistics will be required to reduce the uncertainties on the measured line ratios and fully exploit resonant scattering as an independent constraint on gas velocities in the ICM.

\subsection{Measurements in the context of other clusters and numerical simulations}\label{subsec:comp}

\begin{figure*}
    \centering
    \includegraphics[width=0.5\linewidth]{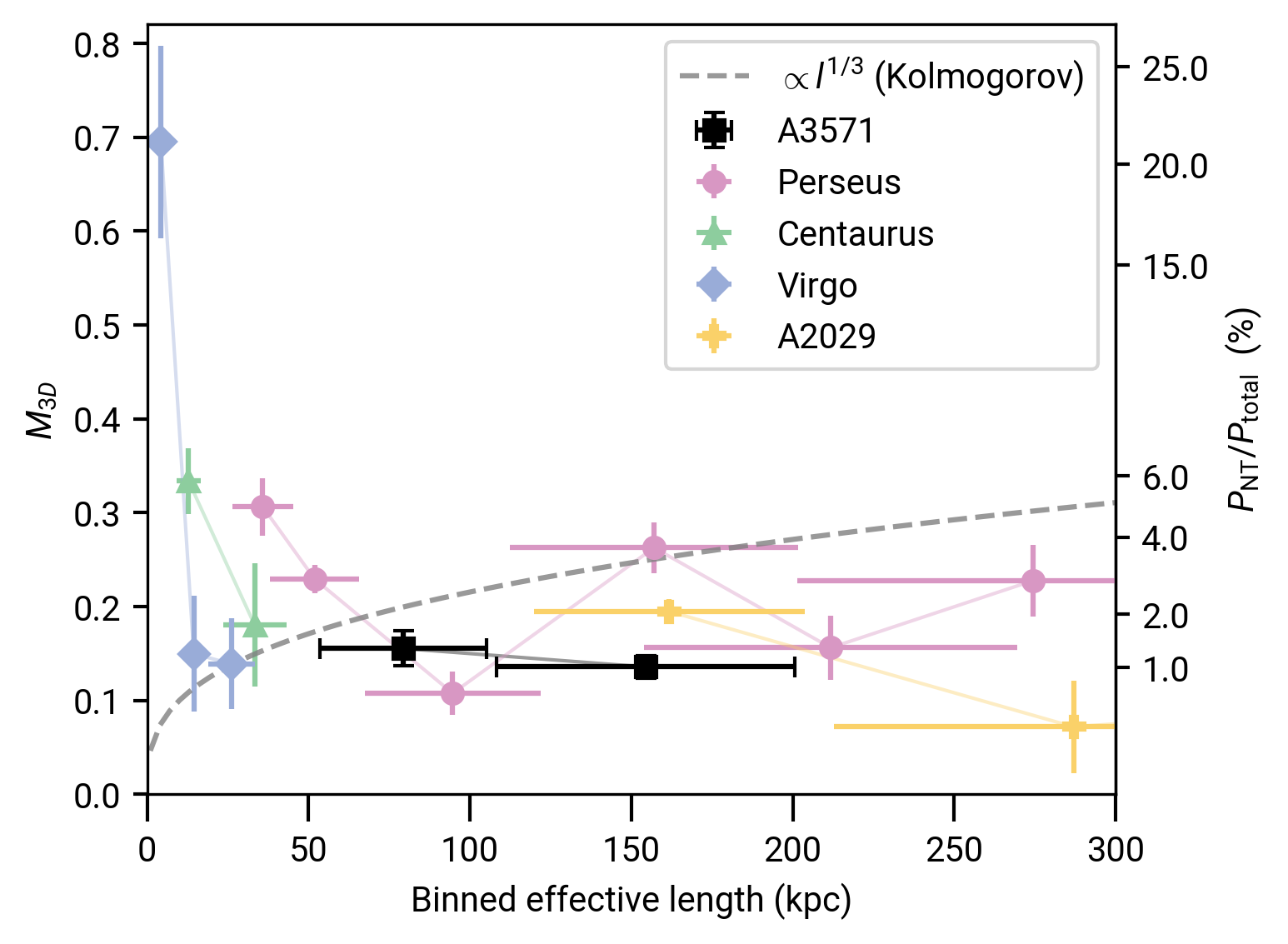}
    \includegraphics[width=0.44\linewidth]{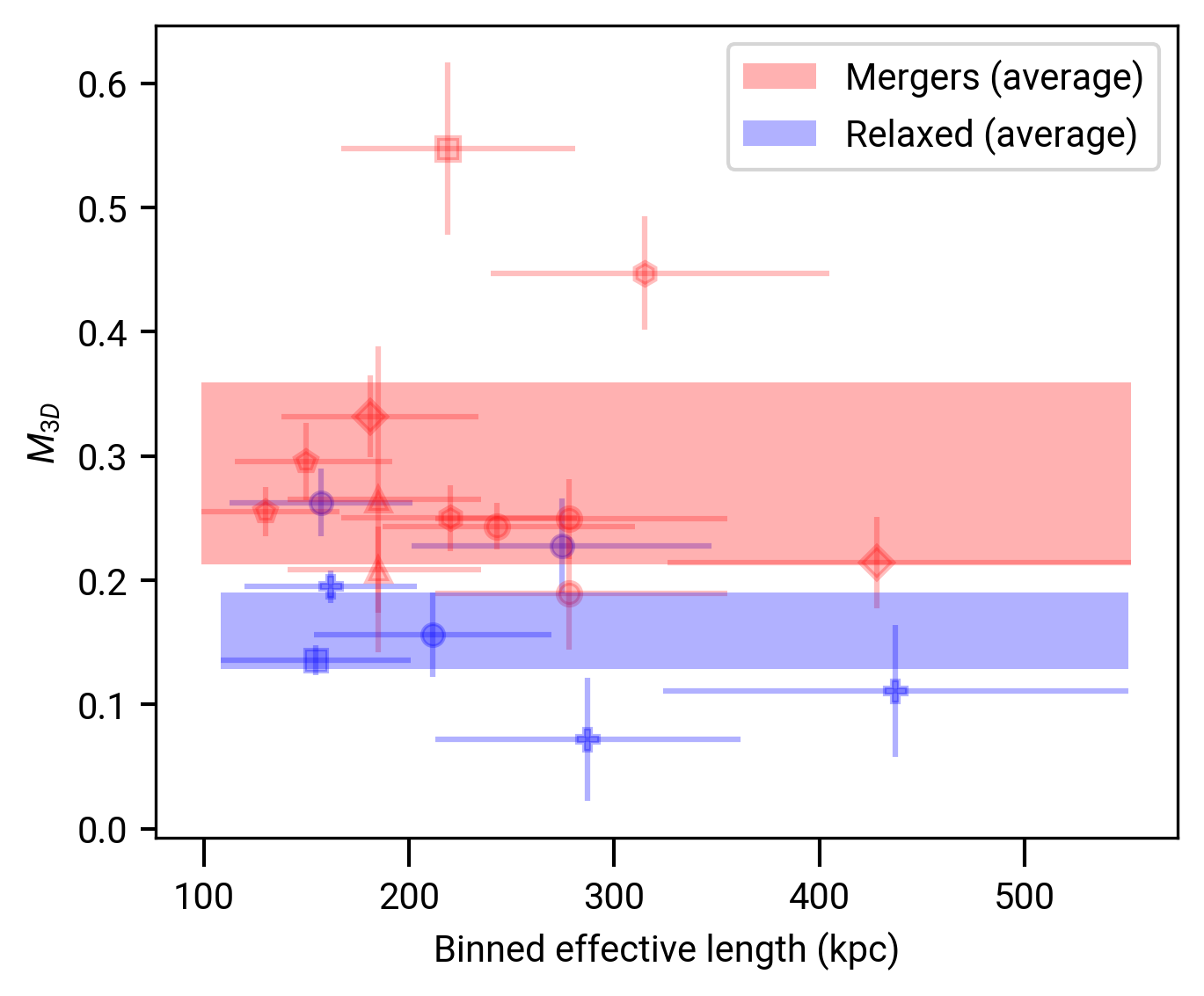}
    \caption{Comparison to other clusters. Left, 3D Mach number (left y-axis) and non-thermal pressure fraction (right y-axis) of A3571 shown with a small sample of clusters for which radial information is available. Right: Average of mergers (A2319, A754, A1914, A2034, A3667, Coma) and relaxed clusters (A2029, A3571, Perseus beyond 100 kpc) in the effective length ranges $99-552$\,kpc and $108-550$\,kpc, respectively, with each sample shown as background points.}
    \label{fig:comparison}
\end{figure*}

XRISM has observed dozens of galaxy clusters in the nearly three years since its launch \citep[e.g.,][]{XRISM_Perseus2026, M87_paper1, XRISM_A2319, XRISM_A2029core, XRISM_Centaurus}. Early observations primarily targeted bright, AGN-active cool-cores, but more recent observations have included a wider sample of merging/dynamically disturbed and relaxed AGN-inactive clusters \citep[e.g.,][]{Fujita_2025, Heinrich_2026, Omiya_2026}. The resulting gradually compiled sample spans a broad range of dynamical and thermodynamical states, providing an opportunity to compare ICM properties across different evolutionary stages.

Such comparisons, however, should be done with caution, accounting for the three-dimensional structure of galaxy clusters. 
Observations measure emissivity-weighted, projected-along-the-line-of-sight gas characteristics. Therefore, quantities within a projected radial bin originate from gas distributed over a range of projected radii and depths along the line-of-sight within the cluster. Consequently, the characteristic spatial scale (or radial bin) associated with a measurement is not necessarily equivalent to the size of the projected radial bin, particularly in cases where the X-ray emissivity is centrally peaked. Determining an appropriate line-of-sight length scale is therefore important when comparing thermodynamic or kinematic quantities between clusters.

To account for this effect, we calculate the distance along the line of sight enclosing 50\% of the total X-ray flux contributing to a given projected radial bin, or the effective length $l_{\mathrm{eff}}$, following \cite{Zhuravleva2012}. We estimate the uncertainties on $l_{\mathrm{eff}}$ using the scales enclosing 40\% and 60\% of the total flux. Using a spherically-symmetric Chandra deprojected density profile (Fig.~\ref{fig:radprofs}), we find $l_{\mathrm{eff}}$ values of $79^{+29}_{-22}$ kpc and $157^{+52}_{-40}$ kpc for R1 and R2, respectively. The effective length as a function of projected radius is shown in the left panel of Figure~\ref{fig:efflength}. For the purpose of our heating and cooling calculations in Section~\ref{subsec:heat}, we also estimate the effective length for the four-pixel region used in \cite{Aihara2026}. Assuming the center of their four-pixel region is the center of the cluster, we find $l_{\mathrm{eff}}=66^{+27}_{-20}$.

We follow the same prescription to calculate effective lengths for a sample of clusters with published central ICM measurements and radial information \citep{XRISM_Perseus2026, M87_paper1, XRISM_Centaurus, XRISM_A2029outskirts}, with the goal of comparing the kinematic properties of clusters in different stages to those of A3571 on comparable physical scales. To account for differences in ICM temperature, and thus sound speed, we calculate $M_{\mathrm{3D}}$ for each cluster using published velocity dispersions and temperatures from XRISM observations. The left panel of Figure~\ref{fig:comparison} shows $M_{\mathrm{3D}}$ as a function of $l_{\mathrm{eff}}$ for the selection of clusters, including the AGN-active cool-core systems Virgo (blue), Perseus (pink), and Centaurus (green), as well as the relaxed cluster A2029 (yellow). A3571 measurements (black points) are shown for comparison.

As illustrated in Figure~\ref{fig:comparison}, the AGN-active cool-core clusters exhibit a pronounced rise in $M_{\rm 3D}$ toward smaller effective lengths ($\leq 50$\,kpc), reflecting enhanced central turbulent motions. A3571 probes $>50$\,kpc scales, preventing a direct comparison of core kinematics on identical physical scales. The effective lengths probed in A3571 are more comparable to those in the sloshing-dominated regions of Perseus, and to the core of A2029, a relaxed cluster exhibiting a weak sloshing spiral structure. On these scales, A3571 displays lower than average values of $M_{\mathrm{3D}}$ for relaxed systems, which range $\sim0.1-0.3$, as seen in Figure~\ref{fig:comparison}.

The non-thermal pressure fraction in A3571 is $1.3\pm0.2\%$ and $1.0\pm0.2\%$ in the R1 and R2 regions, respectively. Region D2 of the default binning, for which we have only an upper limit on velocity dispersion, corresponds to $P_{\mathrm{NT}}/P_{\mathrm{tot}}<0.4\%$. As can be seen in the left panel of Figure~\ref{fig:comparison}, right y-axis, the non-thermal pressure fractions of A3571 are among the lowest in relaxed systems on similar scales. A3571 measurements even fall below the core of the famously-relaxed A2029 \citep{XRISM_A2029core}, which exhibits $\sim2\%$. 

Unlike AGN-active cool cores, which exhibit sharply peaked X-ray emissivity profiles and correspondingly small $l_{\mathrm{eff}}$, merging systems generally probe larger $l_{\mathrm{eff}}$, comparable to those observed in A3571. To compare A3571 and other relaxed systems (Perseus and A2029) directly with merging systems on similar physical scales, we compile a sample of mergers from published measurements for A2319 \citep{XRISM_A2319}, A754 \citep{Omiya_2026_754}, A1914 \citep{Heinrich_2025}, A2034 \citep{Heinrich_2026}, A3667 \citep{Omiya_2026}, and Coma \citep{XRISM_Coma}. Mach number as a function of the binned effective length for these systems, which falls in the range $l_{\mathrm{eff}}=99-552$\,kpc, is shown in the right panel of Figure~\ref{fig:efflength}. We calculate the uncertainty-weighted average $M_{\mathrm{3D}}$ using all measurements with $l_{\mathrm{eff}}$ in this interval for the merging sample, and do the same for the relaxed sample across a similar scale range of $l_{\mathrm{eff}}=108-550$\,kpc.  Uncertainties are estimated using the 16th and 84th percentiles of the distributions. The right panel of Figure~\ref{fig:comparison} shows a clear separation between the two populations, with merging clusters exhibiting a higher average Mach number of $0.29\pm0.07$ (red) compared to $0.16\pm0.03$ for relaxed systems (blue). 

We compare these observational results to predictions from the Omega500 non-radiative cosmological simulations \citep{Nagai2007a, Nagai2007b, Nelson2014a} presented in \cite{Zhuravleva2023}. That work provides radial profiles of the one-component Mach number for subsamples of relaxed, intermediate, and unrelaxed clusters. The smallest radius probed in the simulations was $\sim0.25r_{500c}$, whereas our observational averages span approximately $0.05-0.3r_{500c}$. We therefore adopt the simulated values at the smallest available radius for comparison to our measurements. We note, however, that the simulation predictions are averaged in radial shells and not emissivity-weighted. Consequently, we do not expect exact agreement between simulations and observations; rather, the comparison is intended to assess whether overall trends between relaxed and unrelaxed systems are broadly consistent. We therefore compare the ratio of $M_{\mathrm{3D}}$ in merging systems to relaxed systems for the Omega500 simulations and XRISM observations, finding $2.6\pm1.0$ for simulations and $1.8\pm0.6$ for observations. 
The observed and simulated ratios are consistent within the uncertainties, suggesting that the enhancement of turbulent motions in merging systems as compared to relaxed systems is broadly reproduced by the simulations.

We also compare A3571 to the cosmological simulations examined by \cite{XRISM_sims}, who compared observed cluster velocity dispersions with predictions from the TNG-Cluster, GADGET-X, and GIZMO-SIMBA simulation suites. That study investigated both cool-core and non-cool-core systems to test whether current feedback prescriptions reproduce the observed level of ICM turbulence. The authors found that, on average, simulations overpredict velocity dispersions in cool cores, while showing better agreement with non-cool-core systems. 

Although A3571 is often classified as a weak-cool-core or intermediate system, the R1 region satisfies the cool-core criterion adopted in that work, that the XRISM pointing lies within the radius where the cooling time equals the Universe's age at the cluster redshift. We therefore compare the R1 measurement to the simulated cool-core sample. Because the simulations were divided into mass bins matched to the observational sample, we compare A3571 to analogues of PKS 0745-191, whose mass ($\log_{10}(M_{200c}/M_{\odot})=14.8$) is most similar. PKS 0745-191 is an AGN-active cool-core cluster exhibiting clear X-ray cavities, yet displays velocity dispersion ($\sigma_v=121\pm17~\mathrm{km~s^{-1}}$) significantly lower than values predicted by simulations ($\gtrsim 200~\mathrm{km~s^{-1}}$). Since A3571 shows a similar central $\sigma_v$ to PKS 0745, it also falls below the simulated cool-core population, providing another observational example at odds with predictions from cosmological simulations. The existence of such systems suggests that the treatment of ICM physics in simulations may be inadequate, whether due to missing physical processes or to the specific implementations adopted for those processes. We note, however, that \cite{XRISM_sims} used full XRISM pointings rather than subdivided radial regions as adopted here (though the low variance between R1 and R2 suggests a similar value for the full pointing), and that differences in effective line-of-sight scales are not accounted for in this comparison.

\subsection{Heating and cooling in A3571}\label{subsec:heat}

\begin{figure}
    \centering
    \includegraphics[width=0.95\linewidth]{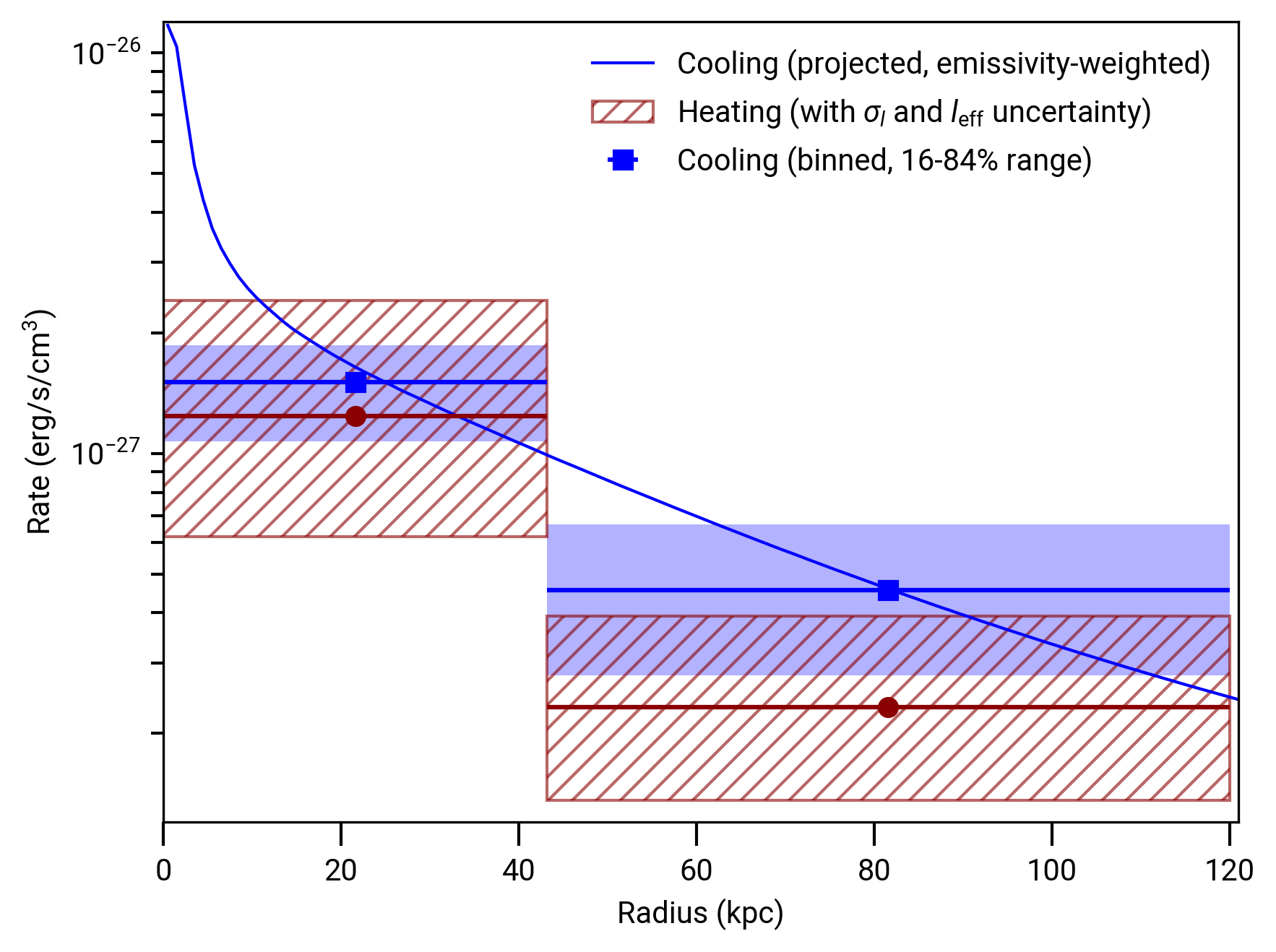}
    \caption{Heating and cooling rates as a function of projected radius. The binned cooling (blue squares) was derived from the projected emissivity-weighted cooling profile (blue curve) by averaging in the same radial bins used for velocity dispersion measurements. The quoted uncertainties (shaded blue regions) on cooling represent the 16-84\% values within each bin and quantify the spatial variation of the projected cooling field across the bin. The heating rate (red circles) is computed from the measured velocity dispersion, derived effective length, and projected, emissivity-weighted density profile. The uncertainties (red hatched regions) reflect the combined  uncertainties on these values.}
    \label{fig:heatcool}
\end{figure}

With direct velocity measurements from XRISM/Resolve, one can estimate the turbulent heating rate for comparison to the radiative cooling losses to assess whether turbulent dissipation may be a significant heat source. In this analysis, we assume that the measured gas motions correspond to volume-filling turbulence. This assumption may not hold if motions are intermittent or include coherent flows \citep{XRISM_Coma, Eckert_2025}. 
In particular, the early-stage sloshing motions observed in A3571 may contribute to the measured velocity dispersion, even though such coherent motions do not necessarily correspond to turbulence. Additionally, since A3571 is a sloshing cluster whose motions are assumed to be predominantly in the plane of the sky \citep{Aihara2026}, the line-of-sight velocities could substantially underestimate the true three-dimensional velocity.
Nevertheless, under the assumption that our measurements correspond to turbulence, we calculate the heating rate via
\begin{equation}\label{eqn:heat}
Q_{\rm heat} = C_0 \rho(r) \sigma_l^3/l
\end{equation} 
where $C_0= 3^{3/2} 2\pi / (2 C_K)^{3/2}$, $C_k=1.65$ is the Kolmogorov constant, $\rho=(n_i+n_e)\mu m_p$ is the gas mass density, and $\sigma_l$ is the velocity amplitude at a given spatial scale $l$. The number densities of electrons $n_e$ and ions $n_i$ are related as $n_i=(\xi-1)n_e$, where $\xi=1.912$ for a fully ionized plasma. This approach explicitly assumes a Kolmogorov turbulent power spectrum \citep{Kolmogorov1941}, and requires caution when selecting the length scale $l$. We take $l$ to be the average effective length $l_{\mathrm{eff},50\%}$ in each radial bin and assume that $\sigma_v$ is dominated by the velocity at this largest probed scale (i.e., $\sigma_l = \sigma_v$), as expected for a turbulent cascade when the injection scale exceeds $l_{\mathrm{eff}}$. For the calculation of $\rho$, $n_e$ is projected and emissivity weighted. Heating is represented by red points in Figure~\ref{fig:heatcool}. Uncertainties, shown via hatched regions, include both the measured uncertainties on $\sigma_v$ and the 40-60\% scales of $l_{\mathrm{eff}}$.

We compare the heating rate to the projected emissivity-weighted cooling rate, which is derived from the 3D cooling rate $Q_\mathrm{cool, 3D}$:
\begin{equation}
    Q_{\rm cool, 3D}=A_0\rho^2\Lambda(T)
\end{equation}
where $A_0$ is the cooling rate coefficient given by $A_0=(\xi-1)/(\xi\mu m_p)^2$. $Q_\mathrm{cool, 3D}$ was projected along the line of sight and evaluated on a uniform grid; it is shown as a blue curve in Figure~\ref{fig:heatcool}. The resulting 2D field was then azimuthally averaged in annuli defined by the region edges of R1 and R2. Within each annulus, the cooling rate was computed as the mean of all pixels, plotted as blue points. The associated uncertainty (blue shaded regions) was estimated from the 16th and 84th percentiles of the pixel distribution within each annulus, providing a measure of the spatial variation of the projected cooling rate across the bin.

Although both heating and cooling rates have large uncertainties on the order of $\sim0.5$ dex, Figure~\ref{fig:heatcool} shows that the estimates are broadly in agreement, signifying that our calculated turbulent dissipation rate is consistent with maintaining the thermal balance of the system. There is no evidence for either runaway cooling or strong overheating in the regions probed by our measurements. 

If the central region of the cluster is instead defined as the four-pixel region used in \cite{Aihara2026}, rather than the nine-pixel region R1, the upper limit on velocity dispersion leads to an upper limit on the heating rate of $3.2\times 10^{-28}\ \mathrm{erg\, s^{-1} cm^{-3}}$ within a projected radius of $25$\,kpc. This calculation assumes that the center of the four-pixel region coincides with the center of the cluster, while there is in fact a $\sim 15$\,kpc offset between the two. In this case, the upper limit on the heating rate falls below the cooling rate within the same region. However, over the $\sim25-43$\,kpc region, the heating and cooling rates are again consistent within the uncertainties. The average of these local heating and cooling rates within $\sim43$\,kpc is consistent with the results obtained using the nine-pixel binning strategy.

The comparison between these spatial scales suggests that the apparent balance between heating and cooling may not be uniform throughout the central region. The four-pixel region probes smaller spatial scales, while R1 averages over a larger region and is more closely centered on the X-ray peak and BCG. In addition, the smaller region has a higher fraction of photons originating from surrounding regions ($\sim50\%$ vs $30\%$), making the inferred heating rate more susceptible to uncertainties in the SSM modeling. Nevertheless, the four-pixel result illustrates that local departures from the overall heating-cooling balance may be present; heating may fall below cooling in the central $25\,$kpc, while balancing on larger scales.

There are several possible sources of heat that could contribute to the approximate thermal balance observed on larger scales. If A3571 is, as suggested in previous works, a late- or post-merger cluster, it could be that bubbles from previous AGN outbursts responsible for supplying heat were disrupted by sloshing, leaving no visual evidence of their previous existence. An alternative explanation, more consistent with the scenario of an early merging stage suggested by \cite{Aihara2026}, is that sloshing and turbulent motions resulting from the merger have so far supplied enough heat to largely counteract the short central cooling times \cite[$1.1^{+0.4}_{-0.3}$ Gyr within 4.2 kpc, $7.7$\,Gyr within 55 kpc,][]{Birzan2012}, obviating the need for AGN feedback. Simulations of idealized mergers and subsequent sloshing found that sloshing could prevent the build-up of cool gas in the core for $\sim 1$ Gyr after minor mergers and several Gyr following major disturbances \citep{ZuHone2010}. \cite{Aihara2026} identified a plausible merger configuration for A3571 that captures the observational features of the cluster; in that simulated merger scenario, A3571 was perturbed by A3572, a nearby galaxy concentration, approximately 1 Gyr ago. In this case, it is plausible that sloshing is the heat source counteracting radiative cooling, and that A3571 is in an early stage of merging, rather than a late one as previously suggested.

However, the apparent balance between heating and cooling on the scales probed here does not necessitate thermal balance on smaller scales. Our R1 measurement probes scales down to $\sim40\,$kpc, whereas the smaller region considered by \cite{Aihara2026} extends to only $25$\,kpc and yields an upper limit on heating below the cooling rate. The larger-scale balance could thus coexist with a local deficit of heating in the innermost region. Sloshing could have provided sufficient heating to offset cooling, on average, across the central region since the merging interaction, while now, the cooling rate dominates heating in the innermost region.

The approximate balance of cooling and heating extends to the R2 bin ($\sim40-120$\,kpc). XRISM observations of the Perseus Cluster \citep{XRISM_Perseus2026} indicate that AGN-driven motions dominate the kinematics within $\sim60$\,kpc, while at larger radii the velocity field is increasingly influenced by sloshing and merger-related processes. In their examination of cooling and heating, \cite{XRISM_Perseus2026} found that these rates were comparable in most regions out to $\sim250$\,kpc, with the possible exception of a region spanning $\sim50-100$\,kpc, where predicted turbulent heating falls an order of magnitude below the radiative cooling. The presence of approximate balance at similar radii in A3571 is therefore not unprecedented; however, unlike Perseus, where both AGN- and merger-driven processes contribute to the heating budget, current data of A3571 shows no clear evidence of ongoing AGN feedback activity. This suggests that, in this system, merger-induced motions may play a more dominant role in regulating the thermodynamic state of the ICM on the resolved scales at these radii. A3571 may represent a phase in the evolution of clusters where merger-induced turbulence replaces AGN-driven turbulence as the dominant heating channel, approximately balancing radiative cooling.

\begin{figure}
    \centering
    \includegraphics[width=0.95\linewidth]{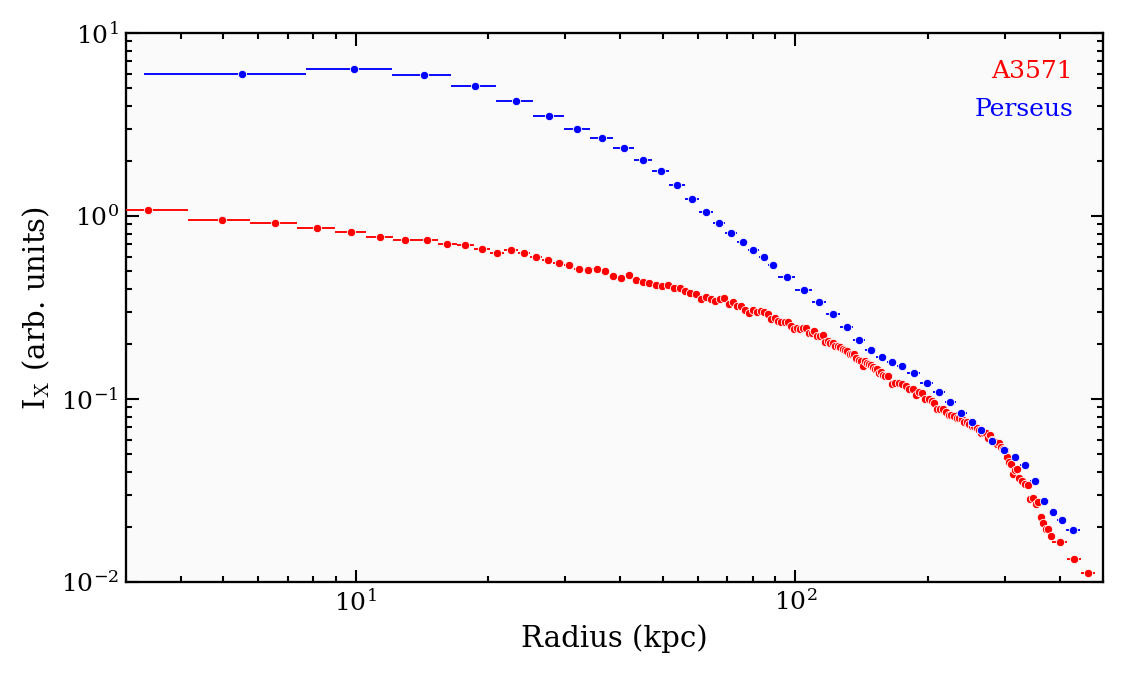}
    \caption{X-ray surface brightness radial profiles ($0.5-8$ keV) for A3571 (red) and Perseus (blue) clusters. A factor of $\sim 7$ difference between the clusters is apparent in the central region.}
    \label{fig:SB}
\end{figure}

Both Perseus and A3571 are $kT \sim 6-7\;\rm keV$ clusters and appear similar in the outskirts, but differ dramatically in the strength of their central cooling flows. Figure~\ref{fig:SB} compares the 0.5--8~keV X-ray surface brightness of these clusters as a function of distance from the core. Inside the central 5--30~kpc, where XRISM results indicate an approximate balance between cooling and heating in Perseus \citep{XRISM_Perseus2026}, the X-ray surface brightness $I_X\propto \rho^2 \Lambda_X(T)R$ in Perseus is $f\sim 6-8$ times higher than in A3571. Since the X-ray emissivity $\Lambda_X(T)$ at these energies does not depend strongly on temperature, the weaker cool core in A3571 implies a gas density $\rho\propto \left ( I_X/R \right )^{1/2}$ smaller than in Perseus by a factor $\sim f^{1/2}$ within $R\sim 30\,\rm kpc$. Assuming that the heating rate $\propto \rho\sigma^3/l$ balances cooling $\propto  \rho^2 \Lambda_{\rm bol}(T)$ in both clusters, we expect 
$
\sigma\propto \rho^{1/3} \left(\Lambda_{\rm bol}(T)l\right )^{1/3}
$
, where $\Lambda_{\rm bol}(T)$ is the bolometric gas cooling function. 
Since A3571 and Perseus have similar temperatures and comparable scales, we expect
$\sigma\propto \rho^{1/3}\sim f^{1/6}$, i.e., a factor of $\sim 1.4$ smaller than that in the Perseus cluster. These numbers broadly agree with our findings, suggesting that the observed level of turbulence in A3571 might approximately compensate for cooling losses or lead to a net heating of the gas in the core (given the uncertainties in the above estimates). However, this could be a ``casual heating'' by sloshing motions, occurring as an incidental by-product rather than a negative feedback loop. If a cool core subsequently develops, energy input from the AGN may become necessary to offset radiative cooling, marking a transition to an AGN-feedback regime. The evidence for the beginning of such a transition may be present in the excess cooling over the upper limit on heating in the $\sim25$\,kpc region.

\section{Conclusion}

In this work, we presented measurements of the weak cool core, AGN-inactive cluster A3571 based on four GO Cycle 1 XRISM pointings, with a particular focus on gas motions in the cluster core. We adopted two different binning strategies: a morphology-motivated division into three regions (D1, D2, D3), and a simpler radial division into two regions (R1, R2). The spectra were modeled using a single-temperature collisionally-ionized thermal plasma model together with treatment of the NXB and SSM between regions. We additionally explored more complex spectral models and a range of potential systematic uncertainties, finding no statistically significant improvement and concluding that the reported measurements are robust.

From the best-fit models, we constrained both the bulk velocity and the velocity dispersion of the ICM. We found a core velocity dispersion of $\sim116~\mathrm{km~s^{-1}}$, with consistent values recovered using both binning strategies. Outside the core, the surrounding regions exhibited similar velocity dispersions of $\sim100-115~\mathrm{km~s^{-1}}$, indicating a relatively uniform velocity amplitude across the scales probed by the observations. The exception is the northern extension of the ICM (region D2), for which only a low upper limit on the velocity dispersion could be obtained. Bulk velocities are modest throughout the regions: the core regions (D1 and R1) display slightly redshifted measurements, while the elongated sloshing region D2 is blueshifted, and the outer regions (D3 and R2) are approximately consistent with zero bulk velocity. 

We also investigated gas motions using resonant scattering as an independent diagnostic complementary to line broadening measurements. Comparing the observed Fe XXV $z/w$ line ratio to tailored Monte Carlo simulations of resonant scattering based on deprojected density, temperature, and abundance profiles, we found that the observed line ratios are consistent with a broad range of Mach numbers within statistical uncertainties, although the central values favor the non-static case (i.e., gas motions in the ICM). These results indicate that A3571 is a promising target for future resonant scattering studies, but that substantially improved photon statistics are required to place constraints on turbulent velocities. 

By calculating the effective line-of-sight length scales contributing to the measurements, we showed that the A3571 measurements probe scales comparable to the center of A2029, the outer sloshing-dominated regions of Perseus, and merging systems, rather than the central AGN-dominated cores. Assuming isotropic gas motions, measurements were used to estimate Mach numbers and non-thermal pressure fractions. We find that the core of A3571 exhibits both a Mach number and a non-thermal pressure fraction that are lower than those in the center of A2029, a cluster previously identified as exceptionally relaxed and for which no analogs were found in cosmological simulation suites \citep{XRISM_sims}. The similarly low values in A3571 suggest that such dynamical quiescence may not be unique, but could instead indicate a population of objects that are underrepresented in current cosmological simulations.

Averaging Mach number over effective length scales corresponding to merging clusters observed with XRISM, we found that merging clusters exhibit an average Mach number of $0.29\pm0.07$, nearly twice the average for the relaxed sample of $0.16\pm0.03$. The ratio of Mach numbers for merging to relaxed clusters is consistent with predictions from non-radiative cosmological simulations.

Assuming that the velocity field in the cluster is dominated by isotropic, volume-filling Kolmogorov turbulence, we compared the turbulent heating rate inferred from measurements of velocity dispersion with the projected emissivity-weighted cooling rate. With our radial binning strategy, we find a heating rate that is significantly lower than that of the AGN-active, strong cool-core Perseus Cluster, but that nevertheless balances the similarly low cooling rate within the substantial uncertainties, both in the inner core of A3571 and in the surrounding region. Calculations using the finer binning approach from \cite{Aihara2026} suggest that local variations from this approximate balance may occur on smaller spatial scales.
We suggest that as there is no evidence for current or past AGN activity and that A3571 is posited to be in an early or late merging state, merger-driven sloshing motions may currently contribute significantly to the heating budget. This is consistent with prior idealized simulations of mergers that found that sloshing-induced turbulence can temporarily offset cooling following a merger event.

We emphasize that the assumption of isotropic Kolmogorov turbulence may not hold. For instance, the observed line broadening could instead be dominated by velocity gradients associated with relatively large-scale motions that have not yet developed into a cascade and therefore do not heat efficiently. However, once a cascade develops through the nonlinearity of the flow, motions with a similar velocity amplitude would dissipate energy at a rate comparable to the current cooling losses. Another potentially important factor is the low ratio of bulk velocity to the velocity dispersion found in A3571. A straightforward interpretation is that the large-scale motions are predominantly in the plane of the sky \citep[e.g.,][]{ZuHone2018}, implying that the true three-dimensional velocities are significantly larger than those inferred from the line-of-sight component. Alternatively, if the characteristic length is small \citep[e.g.,][]{Zhuravleva2012,Zhang2026}, then the line broadening may dominate. These possibilities, and the associated uncertainties, cannot be unambiguously distinguished for a given cluster. This highlights the need for larger samples of clusters, which would enable a fair statistical analysis and a meaningful comparison with simulated samples.

While this work focused on the kinematics in the core of A3571, an accompanying paper applied different binning strategies to the XRISM GO cycle 1 data with the aim of studying the large-scale velocity structure and constraining the evolution of A3571 \citepalias{Aihara2026}. A3571 would be a promising target for future resonant scattering studies, requiring additional observations of its core to better constrain motions. More broadly, the cluster is a useful example of a weak cool-core cluster for future samples, and one that demonstrates XRISM's ability to probe the diverse kinematic states of clusters.

\section*{Acknowledgments}
This work is supported by NASA grants 80NSSC25K7662, 80NSSC25K7693, and 80NSSC25K0143. CZ acknowledges the support of the Czech Science Foundation (GACR) Junior Star grant no.~GM24-10599M. IK was supported by the Simons Foundation via the Simons Investigator Award to A. A. Schekochihin. WF and CJ acknowledge support from the Smithsonian Institution and the Chandra High Resolution Camera Project through NASA contract NAS8-0306.
This research has made use of data obtained from the Chandra Source Catalog, provided by the Chandra X-ray Center (CXC).
The simulations presented in this paper were carried out using the Midway computing cluster provided by the University of Chicago Research Computing Center.

\bibliographystyle{apsrev4-1}

\bibliography{main}

\begin{appendix}

\section{Investigating the effect of flaring}
\label{ap:flare}
To check for possible contamination due to solar flares in the Resolve data, we extracted lightcurves from the concurrent Xtend observations after reprocessing them with \texttt{xapipeline}. A period of flaring is evident below 2 keV during Xtend observation 201095010. Light curves extracted from a source region of radius 4.5\arcmin centered on the cluster in the band 0.5-10 keV with bin size 128 s using \texttt{xselect} show several points beyond 3$\sigma$ from the median count rate of $\sim7.25$ count/s between $10^5$ and $2 \times 10^5$ s into the observation. When a light curve is extracted with the same parameters but instead in a band similar to the Resolve band, 1.7-10 keV, the flaring is no longer evident. 

A lightcurve was then extracted from the Resolve observation in the band 1.7-11 keV, where there was also no evident flaring. To confirm that there was no contamination, spectra were extracted from the entire Resolve FOV---with the exception of pixels 12 and 29--- for (a) the entire exposure and (b) for the time windows that are flare-free in the Xtend 0.5-10 keV lightcurve. Each spectrum was fit in the 2-11 keV band with the same ICM model, \texttt{TBabs*bvapec}, with free parameters temperature, Ar, Ca, Fe, and Ni abundance, redshift, velocity, and normalization. The abundances of He and C were fixed to 1.0 Solar and the others were tied to Fe. All free parameters were in agreement within 1$\sigma$. We also fit the spectra with the described model and an additional \texttt{TBabs*powerlaw} component, to test the scenario in which the flaring presented as a soft excess. Both the best-fit photon index and normalization were in agreement between the fits, and both fits saw a similar statistical improvement (assessed via the Bayesian Information Criterion or BIC) from the ICM model to the ICM + power-law model. From these tests, we concluded that the solar flaring that contaminates the Xtend observation in the soft band does not impact the Resolve data. We therefore did not exclude any of the exposure time when extracting spectra.

\section{X-ray image and deprojection analysis}\label{subsec:models}

\begin{figure}
    \centering
    \includegraphics[width=0.95\linewidth]{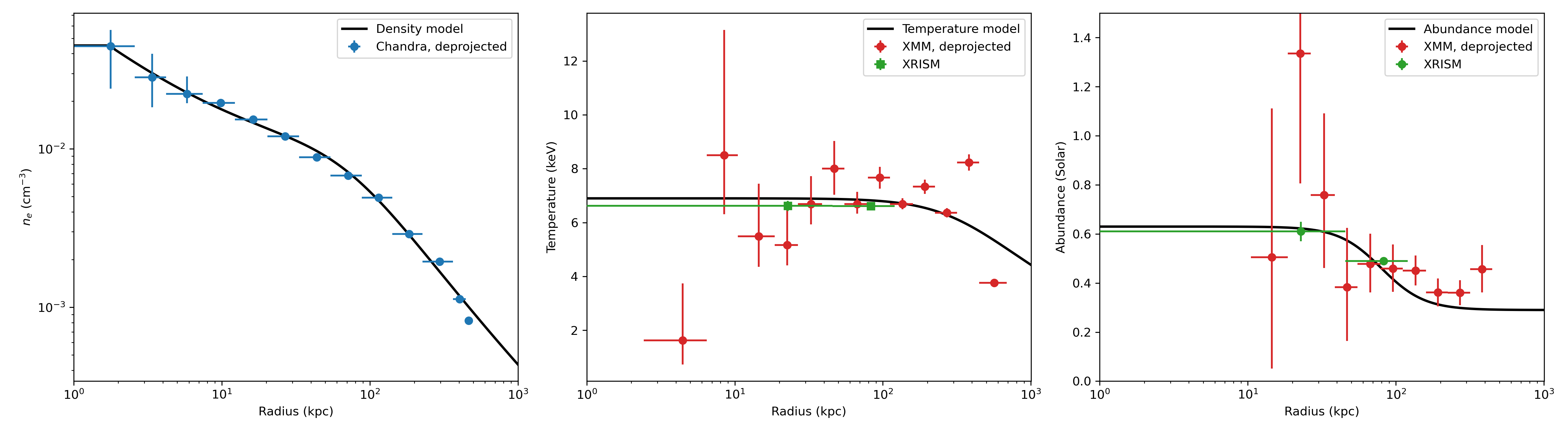}
    \caption{
    Radial profiles and the models for (left) density, (center) temperature, and (right) abundance. The measurements are all deprojected; density is extracted from Chandra data (blue), while temperature and density are from XMM-Newton EPIC/MOS (red). Our projected XRISM/Resolve values are shown in green. The black lines show the best-fit models used in the resonant scattering code.
    }
    \label{fig:radprofs}
\end{figure}

The Monte Carlo simulations of resonant scattering require an input of the deprojected density, temperature, and abundance as a function of radius, while ARF creation requires an input image. In this section, we describe the data reduction of the XMM-Newton and Chandra data required for these steps, and describe our choice of radial models.

A3571 was observed with XMM-Newton for $\sim30$\,ks in July 2002 (ObsID 0086950201). The data were reduced following standard procedures described by \citet{Churazov2003}. Periods affected by soft-proton flares were identified and excluded from the analysis. Blank-sky background files were renormalized to match the observed count rate in the 11–12 keV band. For this study, we used only the EPIC/MOS data.

A3571 was also observed with Chandra for $\sim30$\,ks in July 2003 (ObsID 4203). The data were reprocessed following the standard procedure described by \cite{Vikhlinin2005}. Time intervals affected by flares were removed, and the background treatment accounted for both blank-sky and readout contributions.

To derive radial profiles of the gas density, temperature, and heavy-element abundance, we extracted projected spectra in a series of concentric annuli and deprojected them using the onion-peeling technique described by \cite{Churazov2003}. The resulting deprojected spectra were fitted in the 0.6–8 keV band using XSPEC and the APEC thermal plasma model. We explored several assumptions regarding the metal abundance, including fixing it to a constant value and allowing it to vary freely, to verify that the inferred density and temperature profiles were not sensitive to this choice.

The resulting deprojected profiles are shown in Figure~\ref{fig:radprofs}. For the gas density profile, we relied on the Chandra data because its high spatial resolution provides a more accurate characterization of the innermost few kiloparsecs. The deprojected temperature profile is subject to relatively large uncertainties, owing both to the limited depth of the observations and to the nearly isothermal nature of the cluster core. Nevertheless, the measurements are consistent with the XRISM temperature shown in green. The abundance profile likewise exhibits substantial statistical uncertainties but remains consistent with the XRISM measurements in the central regions.

For the resonant scattering model, we approximated the deprojected profiles with analytical functions. Namely, the density was modeled beyond 1.8\,kpc with a best-fit double-$\beta$ model of the form $n_{\rm e}=n_{\rm 0,1}(1+\left(r/r_{\rm c,1}\right)^2)^{-3\beta_1/2}+n_{\rm 0,2}(1+\left(r/r_{\rm c,2}\right)^2)^{-3\beta_2/2}$, where the normalizations are $n_{\rm 0,1}=0.38$ and $n_{\rm 0,2}=8.6\cdot10^{-3}$ cm$^{-3}$, the core radii are $r_{\rm c,1}=0.084$ and $r_{\rm c,2}=69.97$ kpc, and the slopes are $\beta_1=0.26$ and $\beta_2=0.48$.
We assume a constant value of 0.045 cm$^{-3}$ at radii smaller than 1.8\,kpc. Based on our XRISM measurements, we model the temperature in the core of the cluster as isothermal, and adopt a moderate drop in the outskirts based on the scatter in the XMM-Newton measurements. Our temperature model takes the form $T=6.9(1+(r/350)^2)^{-0.2}$~keV. For the abundance model, we follow the form of the profile described in \cite{Ogorzalek2017}, using the central XRISM measurement as the peak value and a value in the outskirts of 0.3$Z_{\odot}$ \cite[consistent with][]{Werner2013}. The model form is $Z=0.34(2+(r/80)^3)/(1+(r/80)^3)-0.05$.

In addition to the described profiles, we also ran the code with reasonable modifications to the assumed profiles; for example, we tested a steeper and flatter decline of temperature beyond 100\,kpc, modified the normalizations of the profiles by $\pm10\%$, and used a flat abundance profile. All models resulted in $z/w$ ratios within the measurement uncertainties of the XRISM observation.

\section{Additional figures}

\begin{figure}
    \centering
    \includegraphics[width=0.45\linewidth]{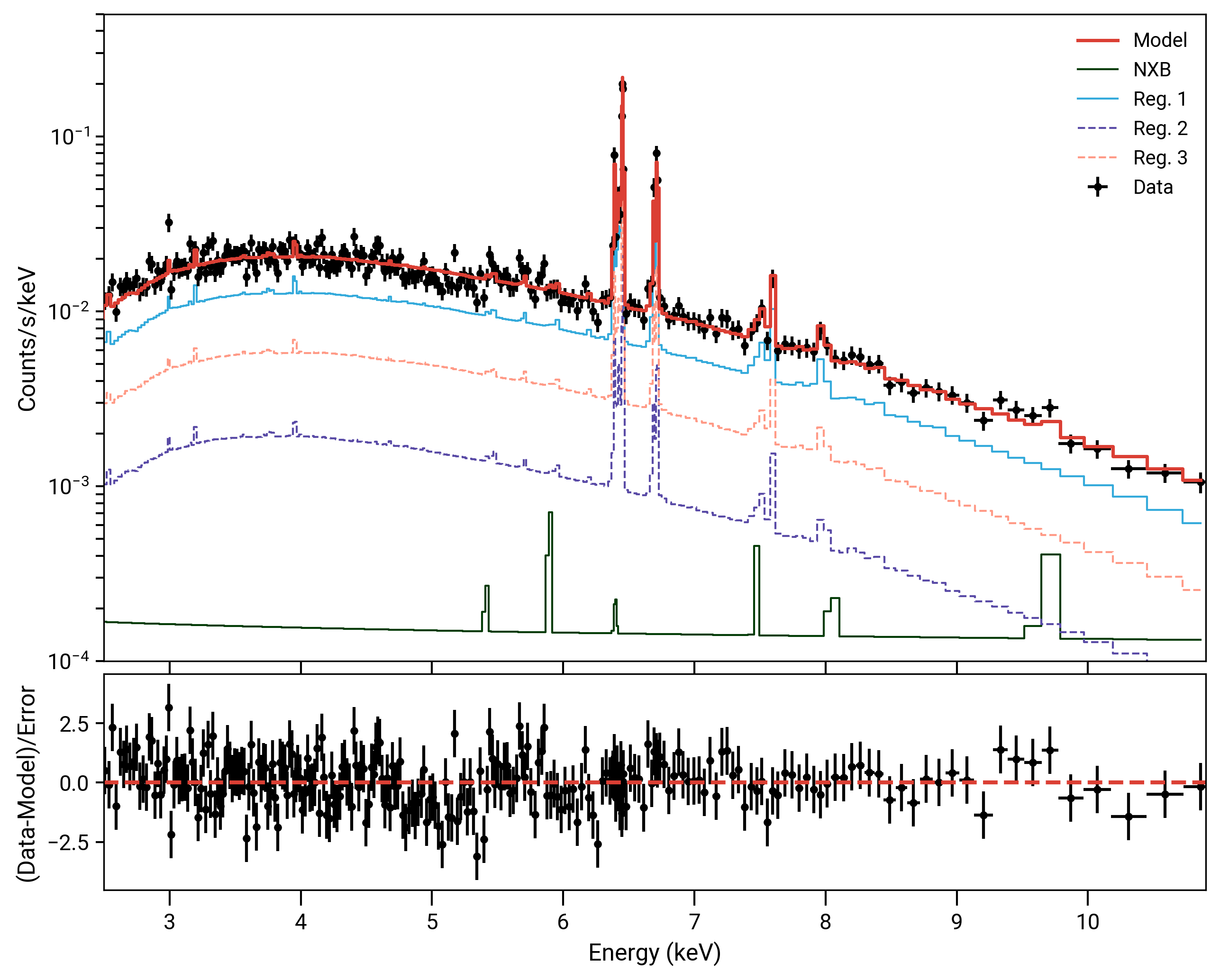}
    \includegraphics[width=0.45\linewidth]{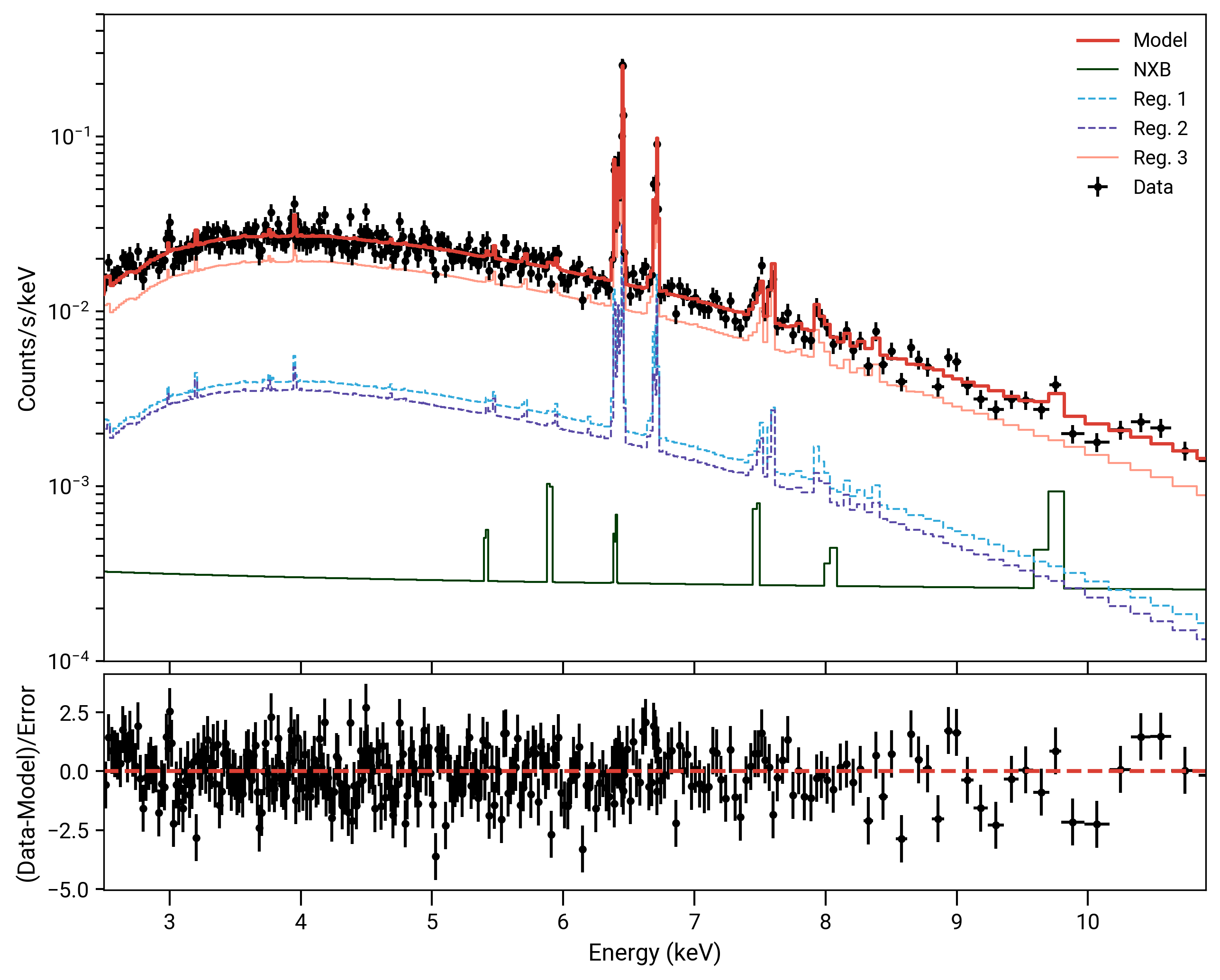}
    \caption{2-11 keV spectrum, best-fit baseline model, and residuals for regions D1 (left) and D3 (right), binned to $>5\sigma$ significance for clear visualization. All components of the fit are shown. SSM-modeled components are shown as dashed lines.}
    \label{fig:broadband}
\end{figure}

The baseline model was fit in the 2-11 keV band, using a single-temperature, collisionally-ionized thermal plasma model affected by Galactic absorption, and accounting for the NXB and SSM. Figure~\ref{fig:spectra} shows the spectra and best-fit model for the default binning strategy around the most prominent lines, but Figure~\ref{fig:broadband} shows the more complete picture, using regions D1 and D3 as examples: the scattered emission contribution shown as dashed lines, the NXB model shown as a solid green line, and the total model in red. The fit residuals are also shown.

\begin{figure}
    \centering
    \includegraphics[width=0.46\linewidth]{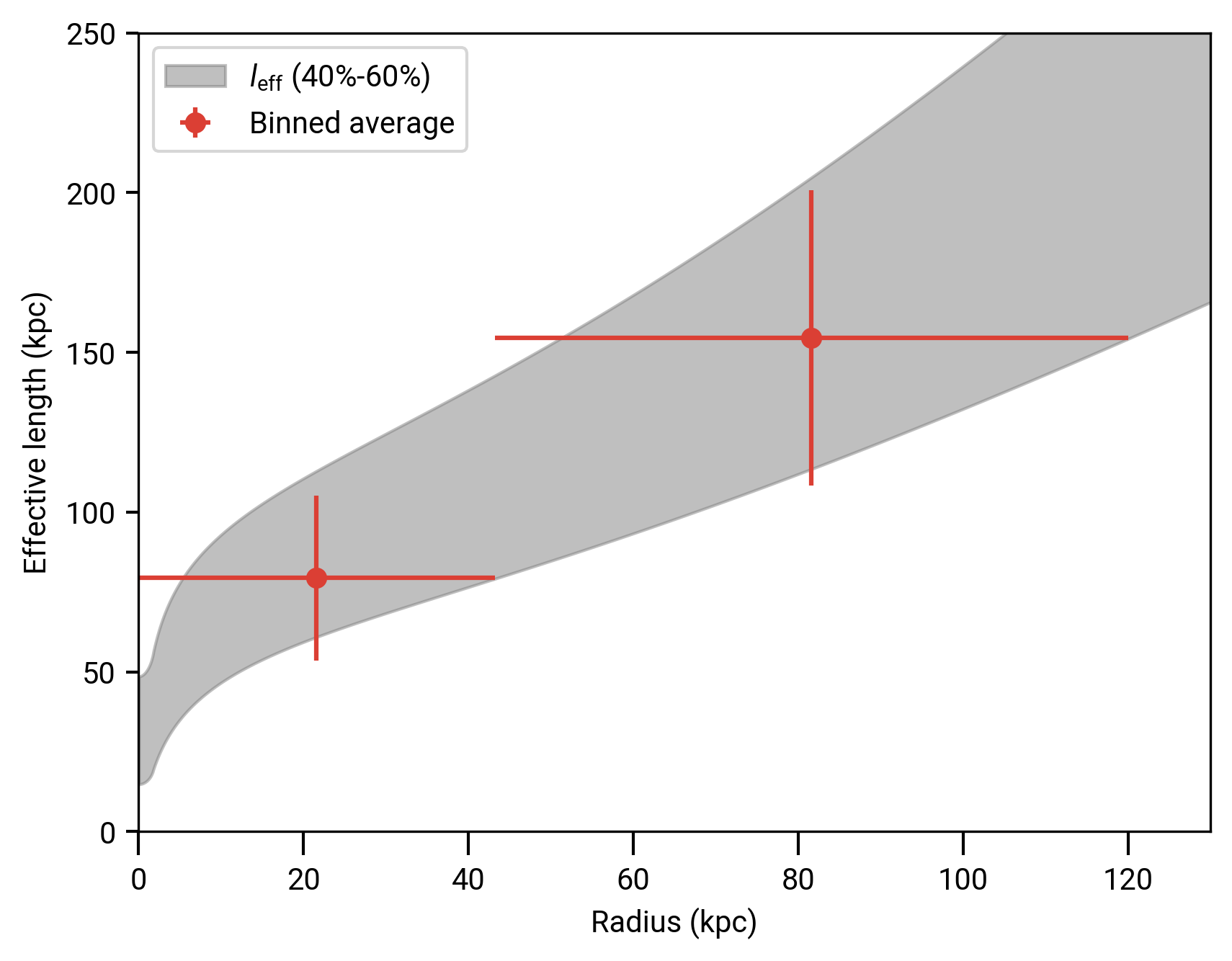}
    \includegraphics[width=0.48\linewidth]{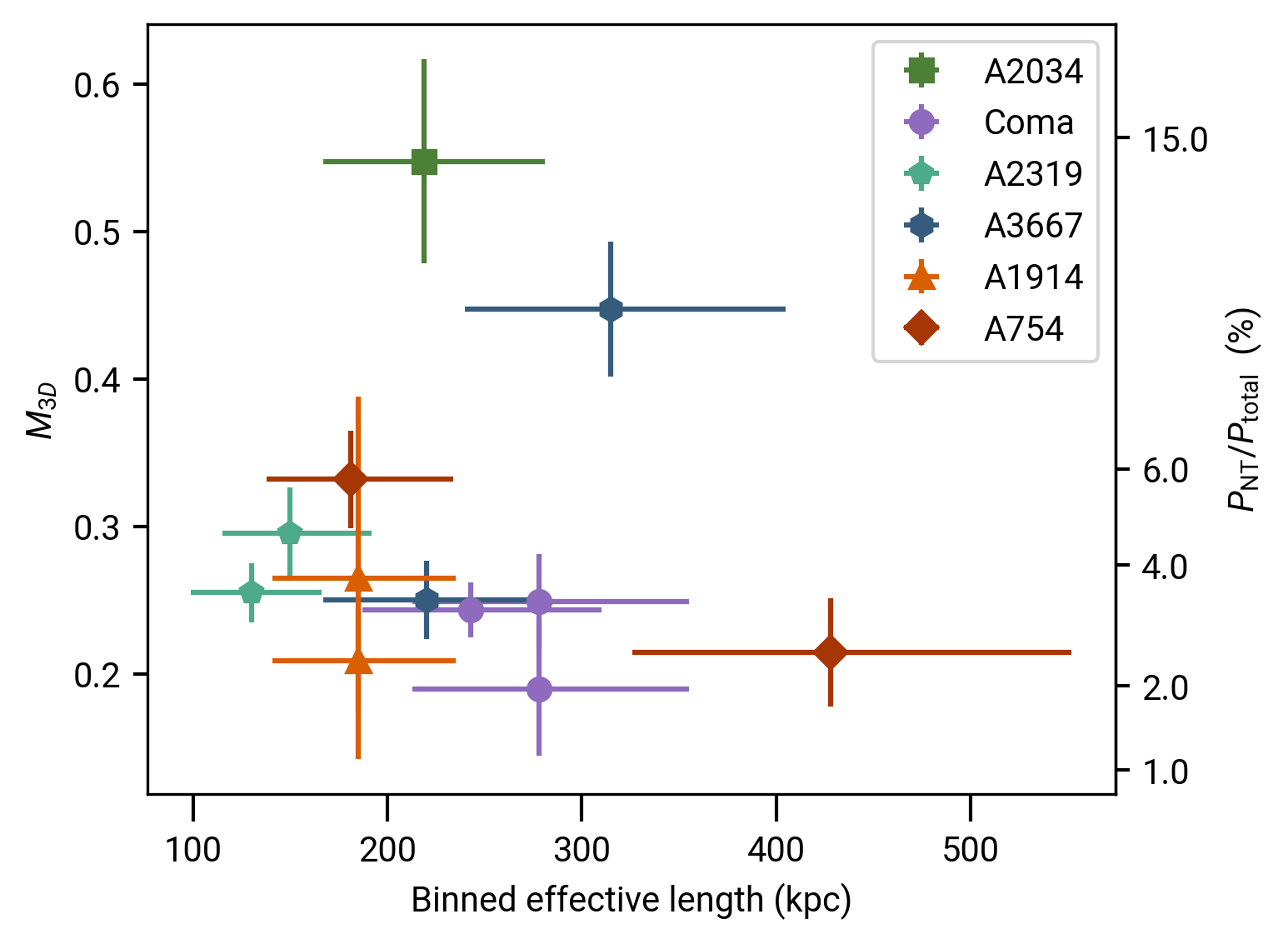}
    \caption{\textit{Left:} Calculated effective length $l_{\mathrm{eff}}$ as a function of projected radius for A3571. The gray shaded region shows the scales enclosing $40-60\%$ of the total flux, while the red data points show the data binned according to the radial binning strategy. \textit{Right:} 3D Mach number (left y-axis) and non-thermal pressure fraction (right y-axis) for the sample of mergers used to calculate the average in Figure~\ref{fig:comparison}.}
    \label{fig:efflength}
\end{figure}

The description of the effective length calculation outlined in subsection~\ref{subsec:comp} describes our choice to calculate the distance along the line of sight enclosing $50\%$ of the total X-ray flux, estimating the uncertainties as enclosing $40\%$ and $60\%$ of the total flux. The resulting effective length for A3571 is shown as a function of the projected radius via a gray shaded region in the left panel of Figure~\ref{fig:efflength}. The red data points present the binned averages for the radial binning strategy.

The right panel of Figure~\ref{fig:efflength} shows the Mach number and non-thermal pressure fraction as a function of effective length for the sample of merging clusters used in subsection~\ref{subsec:comp}. Cluster A1914 has a highly irregular morphology \cite[two merging components along the line of sight, see][]{Heinrich_2025}, making it difficult to model its density profile accurately. Consequently, the uncertainty associated with its effective length is likely underestimated by the reported values.

\end{appendix}

\end{document}